\documentclass[10pt,a4paper]{article}

\usepackage[utf8]{inputenc}
\usepackage[T1]{fontenc}
\usepackage{lmodern}
\usepackage{microtype}
\usepackage{amsmath,amssymb}
\usepackage{graphicx}
\usepackage{hyperref}
\usepackage{authblk}
\usepackage[margin=2cm]{geometry}
\usepackage{booktabs}
\usepackage{tabularx}
\usepackage{pdflscape}
\usepackage{threeparttable}
\usepackage[font=small,labelfont=bf]{caption}
\usepackage{float}
\usepackage{mhchem}

\newcolumntype{L}[1]{>{\raggedright\arraybackslash}p{#1}}

\usepackage[style=numeric-comp,backend=biber,sortcites=true,sorting=none,date=year,maxcitenames=1]{biblatex}

\AtEveryBibitem{%
  \iffieldundef{doi}{}{\clearfield{url}}%
  \clearfield{urldate}%
  \clearfield{urlyear}%
  \clearfield{urlmonth}%
  \clearfield{urlday}%
  \clearfield{issn}%
  \clearfield{eprint}%
}

\newcommand{\textcitenum}[1]{%
  \citeauthor{#1} (\citeyear{#1})~\cite{#1}%
}

\title{REMIND-PyPSA-Eur: Integrating power system flexibility into sector-coupled energy transition pathways}

\author[1,2]{Adrian~Odenweller}
\author[1]{Falko~Ueckerdt}
\author[1,3]{Johannes~Hampp}
\author[1]{Ivan~Ramirez}
\author[1]{Felix~Schreyer}
\author[1,2]{Robin~Hasse}
\author[1,2]{Jarusch~Müßel}
\author[1]{Chen~Chris~Gong}
\author[1]{Robert~Pietzcker}
\author[4]{Tom~Brown}
\author[1,2]{Gunnar~Luderer}

\affil[1]{Potsdam Institute for Climate Impact Research, Potsdam, Germany}
\affil[2]{Global Energy Systems Analysis, Technische Universität Berlin, Berlin, Germany}
\affil[3]{Open Energy Transition, Bayreuth, Germany}
\affil[4]{Department of Digital Transformation in Energy Systems, Technische Universität Berlin, Berlin, Germany}

\date{}

\begin{document}

\maketitle

% Abstract
\begin{abstract}
The rapid expansion of low-cost renewable electricity combined with end-use electrification in transport, industry, and buildings offers a promising path to deep decarbonisation. However, aligning variable supply with demand requires strategies for daily and seasonal balancing. Existing models either lack the wide scope required for long-term transition pathways or the spatio-temporal detail to capture power system variability and flexibility. Here, we combine the complementary strengths of REMIND, a long-term integrated assessment model, and PyPSA-Eur, an hourly energy system model, through a bi-directional, price-based and iterative soft coupling. REMIND provides pathway variables such as sectoral electricity demand, installed capacities, and costs to PyPSA-Eur, which returns optimised operational variables such as capacity factors, storage requirements, and relative prices. After sufficient convergence, this integrated approach jointly optimises long-term investment and short-term operation. We demonstrate the coupling for two Germany-focused scenarios, with and without demand-side flexibility, reaching climate neutrality by 2045. Our results confirm that a sector-coupled energy system with nearly 100\% renewable electricity is technically possible and economically viable. Power system flexibility influences long-term pathways through price differentiation: supply-side market values vary by generation technology, while demand-side prices vary by end-use sector. Flexible electrolysers and smart-charging electric vehicles benefit from below-average prices, whereas less flexible heat pumps face almost twice the average price due to winter peak loads. Without demand-side flexibility, electricity prices increase across all end-users, though battery deployment partially compensates. Our approach therefore fully integrates power system dynamics into multi-decadal energy transition pathways.
\end{abstract}

% Keywords

\section{Introduction}

\subsection{Background}

Limiting global warming in line with the targets of the Paris Agreement requires a fundamental transition of the global energy system towards low-carbon energy sources\parencite{ipccClimateChange20222022,davisNetzeroEmissionsEnergy2018}. With plummeting costs and record growth of variable renewable energy (VRE) sources such as solar photovoltaics (PV) and wind power, renewable electricity is emerging as a central pillar to achieve a deep decarbonisation of the energy system\parencite{ieaWorldEnergyOutlook2024,ludererImpactDecliningRenewable2022}. In most parts of the world, renewables now have a lower levelised cost of electricity (LCOE) than the cheapest fossil fuel alternative\parencite{irenaRenewablePowerGeneration2025}. Simultaneously, rapid cost declines and unprecedented deployment of battery energy storage systems facilitate further integration of VREs\parencite{mallapragadaLongrunSystemValue2020}. As a result, global solar PV generation has doubled within just three years\parencite{emberGlobalElectricityReview2025}. Encouraged by these developments, at the 28th United Nations Climate Change Conference (COP28) the international community formalised the goal to triple global renewable energy capacity by 2030 as part of the UAE Consensus\parencite{unfcccReportConferenceParties2024}, an ambition the International Energy Agency (IEA) has described as the ``single most important lever to bring about the reduction in carbon dioxide (\ce{\ce{CO2}}) emissions needed by 2030''\parencite{ieaTriplingRenewablePower2023}.

To achieve energy system decarbonisation, flexible end-use electrification is critical for cost-effective VRE integration. This increased linking of electricity supply with new electricity demands – referred to as sector coupling\parencite{ramsebnerSectorCouplingConcept2021} – involves meeting energy service demands across the transport, buildings and industry sectors as well as power-to-molecule conversion through renewable electricity. Although electricity currently accounts for only one-fifth of global final energy consumption\parencite{ieaWorldEnergyBalances2024}, virtually all space heating and cooling\parencite{thomassenDecarbonisationEUHeating2021,rosenowHeatingGlobalHeat2022} and the majority of industrial processes\parencite{madedduCO2ReductionPotential2020,rosenowOpportunitiesHeatElectrification2025} can be electrified. With ongoing improvements in battery technologies, road transport is also set for widespread electrification, not just for passengers cars\parencite{hoekstraUnderestimatedPotentialBattery2019} but also for freight transport\parencite{linkRapidlyDecliningCosts2024}. Electricity is therefore positioned to become the dominant energy carrier in a future energy system. This trend becomes even more important as competing emissions abatement options continue to face challenges, exemplified by the sluggish deployment and high costs of carbon capture and storage\parencite{kazlouFeasibleDeploymentCarbon2024} and green hydrogen\parencite{odenwellerGreenHydrogenAmbition2025}, as well as sustainability concerns about large-scale bioenergy use\parencite{jiaLandClimateInteractions2019}. However, sector coupling also leads to considerable operational and planning challenges for future power systems.

\subsection{Challenges for power systems and long-term models}

With increasingly high shares of VRE sources and newly electrified end-use sectors, maintaining the balance between supply and demand at each location and time becomes increasingly challenging\parencite{zhengStrategiesClimateresilientGlobal2025}. Weather-dependent renewable generation creates periods of both surplus and scarcity, including extended low-wind, low-solar periods known as ``Dunkelflaute'' that can last several days or weeks\parencite{kittelMeasuringDunkelflauteHow2024}. However, the electrification of transport through electric vehicles\parencite{schillPowerSystemImpacts2015}, buildings via heat pumps\parencite{bloessPowertoheatRenewableEnergy2018}, industry\parencite{rosenowOpportunitiesHeatElectrification2025} as well as flexible electrolysers\parencite{ruhnauHowFlexibleElectricity2022} also introduces new demand patterns that offer unprecedented opportunities for demand-side management and system flexibility. Successfully harnessing these flexibility potentials, combined with energy storage\parencite{victoriaRoleStorageTechnologies2019,schillElectricityStorageRenewable2020}, transmission grid expansion\parencite{brownSynergiesSectorCoupling2018}, and dispatchable backup capacity\parencite{sepulvedaRoleFirmLowCarbon2018}, is essential to ensure reliable electricity supply while maximising both the integration of VREs and the electrification of end-uses.

\begin{figure}
\centering
\includegraphics[width=\textwidth]{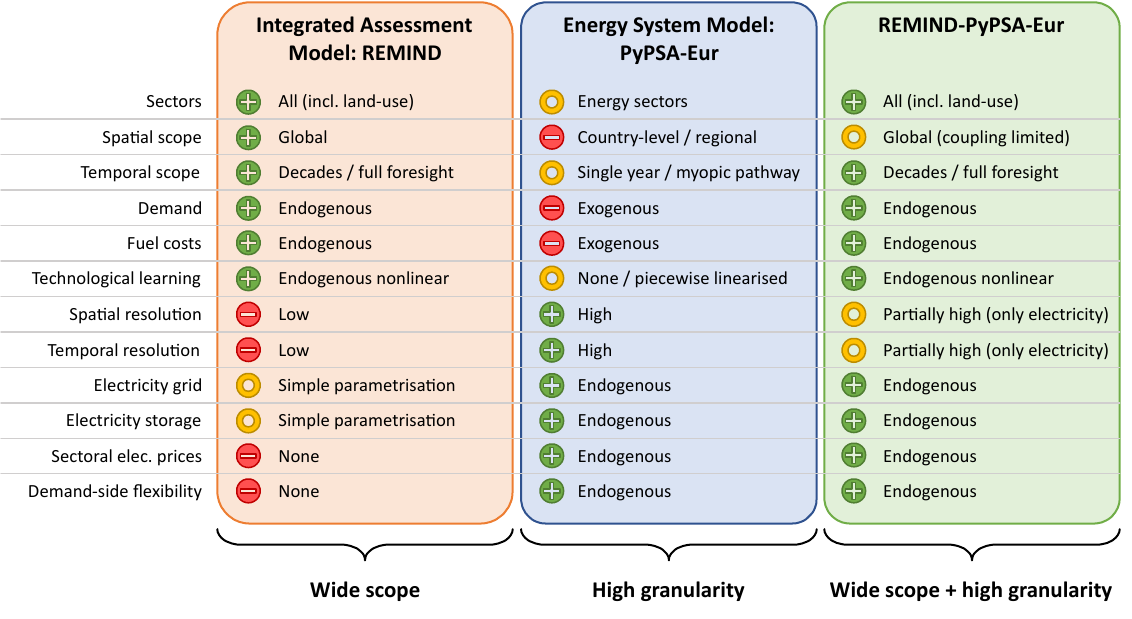}
\caption{\textbf{Complementary strengths of the Integrated Assessment Model REMIND and the Energy System Model PyPSA-Eur (stylised).} REMIND features a wide scope, providing intertemporally optimal transformation pathways over several decades for all economic sectors, but has a low spatio-temporal resolution. PyPSA-Eur features high granularity, enabling a detailed analysis of storage, transmission and infrastructure, but is constrained in temporal and regional scope.}
\label{fig:strengths}
\end{figure}

The challenges and opportunities of flexible power systems are not well represented in long-term integrated assessment models (IAMs) that are regularly used to inform policymakers about national, regional and global energy transition pathways for mitigating climate change. Fundamentally, due to numerical complexity, models cannot have both (i) the wide scope required for cross-sectoral long-term mitigation scenarios (modelled in IAMs) as well as (ii) the high spatio-temporal detail required for power systems operations (modelled in Energy System Models, ESMs)\parencite{ringkjobReviewModellingTools2018}. This creates an inherent trade-off: Models that capture global economic interactions and multi-decadal transition dynamics across all sectors necessarily sacrifice the spatial and temporal resolution required to represent hourly power system dynamics. Vice versa, models that provide high spatio-temporal detail typically focus on shorter time horizons on a regional scope. Yet, developing robust transformation pathways requires bridging these scales in order to combine long-term planning with short-term operations, thereby leveraging the complementary strengths of IAMs and ESMs (Figure~\ref{fig:strengths}).

IAMs like REMIND capture the broad scope and long-term perspective required for climate policy analysis by covering all greenhouse gases including energy, land and carbon dioxide removal (CDR) options and all energy sectors, linked to a representation of the macro-economy, within a global scope. This comprehensive approach enables exploration of cross-sectoral transformation scenarios until the end of the century, subject to different global and regional climate targets and policies. However, as IAMs have a very low spatio-temporal resolution they cannot represent hourly power system dynamics explicitly.

ESMs like PyPSA-Eur feature strengths that are mostly complementary to IAMs, particularly high spatio-temporal detail. This enables endogenous optimisation of dispatch, investment, storage, transmission, and demand-side flexibility for future power systems. However, due to the associated numerical complexity, ESMs typically run at a country or regional level for a single future target year or employ myopic pathway optimisation without the interdecadal foresight essential for long-term planning. Moreover, demand is typically price-inelastic and fuel costs are exogenous.

\subsection{Previous approaches for representing power systems in long-term models}

Various approaches have been developed in recent studies to bridge these scales by enhancing the modelling of short-term power system variations in long-term models. These can be split into (i) approaches that are based on simplified parametrisations of power system dynamics in long-term models and (ii) approaches that establish a soft-link between long-term and short-term models with varying levels of integration. An overview of the strengths and challenges of different approaches is available in \textcite{collinsIntegratingShortTerm2017}.

Early versions of simplified parametrisations often relied on exogenous assumptions for critical variables such as backup capacities\parencite{sullivanImpactsConsideringElectric2013}, with some approaches even imposing hard upper bounds for VREs\parencite{pietzckerSystemIntegrationWind2017}. Alternative approaches used representative days or integration costs, inspired by the system LCOE concept\parencite{ueckerdtSystemLCOEWhat2013}, in order to represent the economic costs of variability\parencite{hirthIntegrationCostsRevisited2015}. A concerted effort within the ADVANCE project in 2017\parencite{ludererAssessmentWindSolar2017} led to the widespread adoption of residual load duration curves (RLDCs) across IAMs. These RLDCs were based on hourly ESM results and parametrised for 8 world regions\parencite{ueckerdtDecarbonizingGlobalPower2017}. Their integration into IAMs led to a dramatic increase of VREs in scenarios, from 38\% to 62\% on average, highlighting the critical role of appropriately representing variability in long-term models\parencite{pietzckerSystemIntegrationWind2017}. More recently, this line of research has been reinvigorated by \textcite{gotskeFirstStepsBridging2025}, who use the sector-coupled PyPSA-Eur model to analyse the effect of imposed VRE shares on key power system metrics, an approach introduced in earlier studies\parencite{ueckerdtDecarbonizingGlobalPower2017,scholzApplicationHighdetailEnergy2017}. However,  these reduced-form approaches only address supply-side variability, but often neglect demand-side flexibility.

In order to address these shortcomings, several studies have contributed towards soft-linking long-term energy models with short-term ESMs, ranging from unidirectional, to manual bidirectional, to automated bidirectional approaches (Table \ref{tab:review}). However, key research gaps remain. First, the majority of approaches are either unidirectional (no feedback) or require manual bidirectional coupling with typically only a single iteration, limiting the ability of fully combining long-term planning with short-term operations. Second, most studies have a limited temporal scope and resolution, coupling only a few selected years and often using rolling horizons instead of full foresight. Third, most studies only use a single aggregated demand profile, precluding the critical analysis of evolving demand patterns from ongoing end-use electrification. Fourth, only very few studies model demand-side flexibility, or limit flexibility to electrolysers. Fifth, all studies except \textcite{gongBidirectionalCouplingLongterm2023} only import selected parameters, often related to backup capacities, from the ESM into the long-term model, whereas full model harmonisation requires a comprehensive parameter exchange. Lastly, so far, no study has incorporated a price-based coupling for both the supply side and the demand side, which is crucial to fully integrate the impact of hourly power system economics into long-term investment decisions.

\begin{landscape}
\begin{table}[p]
\scriptsize
\begin{threeparttable}
\setlength{\tabcolsep}{3pt}
\linespread{0.85}\selectfont
\renewcommand{\arraystretch}{0.75}
\setlength{\aboverulesep}{2pt}
\setlength{\belowrulesep}{2pt}
\caption{Review of soft-coupling approaches between long-term models and short-term energy system models.}
\label{tab:review}

\begin{tabularx}{\linewidth}{L{2.2cm}L{1.6cm}L{1.6cm}L{2cm}XXL{2.8cm}L{1.8cm}L{1.8cm}L{2cm}}
\toprule
& & & & & & \multicolumn{4}{c}{\textbf{Power system modelling}} \\ 
\cmidrule{7-10}
\textbf{Publication} & \textbf{Long-term model} & \textbf{Short-term ESM} & \textbf{Coupling scope} & \textbf{Long-term model to ESM\tnote{a}} & \textbf{ESM to long-term model\tnote{b}} & \textbf{Resolution \& foresight} & \textbf{Electricity storage} & \textbf{Sectoral demand\tnote{c}} & \textbf{Demand flexibility} \\
\midrule
% UNIDIRECTIONAL COUPLING
\multicolumn{10}{l}{\textit{\textbf{Unidirectional coupling}}} \\
\midrule
\textcitenum{deaneSoftlinkingPowerSystems2012} & TIMES & PLEXOS & Ireland; 2020 & Total demand, capacities, costs & - & 1 node; \newline 30-min (1-day RH) & PHS & - & - \\
\cmidrule{1-10}
\textcitenum{deaneAssessingPowerSystem2015} & TIMES (MONET) & PLEXOS & Italy; 2030 & Total demand, capacities, costs, intra-regional trade & - & 6 nodes; hourly (foresight ?) & PHS & - & - \\
\cmidrule{1-10}
\textcitenum{collinsAddingValueEU2017} & PRIMES & PLEXOS & EU-28; 2030 & Total demand, capacities, costs, VRE CFs & - & 28 countries (nodes ?); \newline hourly (foresight ?) & PHS & - & Stylised (10\% peak intra-day) \\
\cmidrule{1-10}
\textcitenum{pavicevicPotentialSectorCoupling2020} & JRC-EU-TIMES & Dispa-SET & EU-28; 2050 & Sectoral demand, capacities, costs & - & 28 nodes; \newline hourly (1-day RH) & PHS, BESS, V2G & EVs, heating & EVs, heating \\
\cmidrule{1-10}
\textcitenum{younisScrutinizingIntermittencyRenewable2022} & TIMES-CO-BBE & PowerPlan (simulation) & Colombia; 2050 & Total demand, capacities, costs & - & 5 zones; hourly (no optimisation) & - & - & - \\
\cmidrule{1-10}
\textcitenum{beresWillHydrogenSynthetic2024} & JRC-EU-TIMES & PLEXOS & EU-27; 2050 & Sectoral demand, costs, max. CCS/bio & - & 27 nodes; \newline hourly (foresight ?) & PHS, BESS, \ce{H2}, CSP & All sectors (incl. \ce{H2}) & \ce{H2} (intra-day) \\
\cmidrule{1-10}
\textcitenum{floresAssessmentImpactsRenewable2024} & GCAM-Chile & \ce{H2}RES & Chile; 2020-2050 & Sectoral demand, capacities, costs & - & 1 node; \newline hourly (foresight ?) & PHS, \ce{H2}, BESS & Heating, \ce{H2} & ? \\
\midrule
% MANUAL BIDIRECTIONAL COUPLING
\multicolumn{10}{l}{\textit{\textbf{Manual bidirectional coupling}}} \\
\midrule
\textcitenum{brinkerinkAssessingGlobalClimate2022} & MESSAGEix-GLOBIOM & PLEXOS-World & Global; 2050 & Total demand, capacities, costs & Inter-regional trade & 258 nodes; mixed resolution \& foresight & Generic 24h storage & - & - \\
\cmidrule{1-10}
\textcitenum{wyrwaNewApproachCoupling2022} & TIMES-PL & MEDUSA & Poland; 2020-2050 (?) & Total demand, capacities, costs & Backup coefficient & 1 node; \newline hourly (2-day RH) & Generic 4h storage & - & Stylised (2GW intra-day) \\
\cmidrule{1-10}
\textcitenum{fernandezvazquezEnergyTransitionImplications2024} & OSeMOSYS & Dispa-SET & Bolivia; 2020-2050 & Total demand, capacities, costs & Backup margin & 4 nodes; \newline hourly (4-day RH) & - & - & - \\
\cmidrule{1-10}
\textcitenum{kleanthisBidirectionalSoftlinkingCapacity2025} & OSeMOSYS & FlexTool & Greece; 2030, 2040, 2050 & Total demand, capacities, costs, VRE CFs, imports & Additional VRE \& BESS capacity (2050 only) & 1 node (?); \newline hourly (full foresight) & PHS, BESS, \ce{H2} & - & - \\
% AUTOMATED BIDIRECTIONAL COUPLING
\midrule
\multicolumn{10}{l}{\textit{\textbf{Automated bidirectional coupling}}} \\
\midrule
\textcitenum{pinaHighresolutionModelingFramework2013} & TIMES & EnergyPLAN & Portugal; 2005-2050 & Capacities & Capacity limit & 1 node (?); \newline hourly (foresight ?) & PHS & - & - \\
\cmidrule{1-10}
\textcitenum{despresStorageFlexibilityOption2017} & POLES & EUCAD & Europe; 2000-2100 & Load curves, capacities, costs & Storage, trade, VRE CFs & 24 nodes; 12 typical days (1-day foresight) & PHS, BESS, CAES, V2G & EVs, \ce{H2} (?) & EVs, \ce{H2} (?) \\
\cmidrule{1-10}
\textcitenum{alimouAssessingSecurityElectricity2020} & TIMES & ANTARES & France; 2030 & Total demand, capacities & Backup (capacity credit) & 1 node; \newline hourly (1-week RH) & PHS & - & - \\
\cmidrule{1-10}
\textcitenum{seljomBidirectionalLinkageLongterm2020} & TIMES & EMPS & Norway; 2030, 2050 & Total demand, capacities & Hydro availability, trade prices & 11 nodes; \newline 2-4 hourly & Hydro / PHS & - & - \\
\cmidrule{1-10}
\textcitenum{gongBidirectionalCouplingLongterm2023} & REMIND & DIETER & Germany; 2020-2100 & Sectoral demand, capacities, costs, VRE CFs & Backup, storage, non-VRE CFs, market values & 1 node; \newline hourly (full foresight) & BESS, \ce{H2} & \ce{H2} & \ce{H2} \\
\cmidrule{1-10}
\textcitenum{rosendalBenefitsChallengesSoftlinking2025} & Balmorel & ANTARES & Europe; 2020-2050 & Capacities, costs & Backup (different adequacy metrics) & 52 nodes; \newline hourly (1-week RH) & BESS, \ce{H2} & \ce{H2} & \ce{H2} \\
\cmidrule{1-10}
\textbf{This study} & REMIND & PyPSA-Eur & Germany; 2030-2100 & Sectoral demand, capacities, costs, EV fleet size & Backup, all CFs, storage, grid losses, market values, sectoral prices & 4 nodes, \newline hourly (full foresight) & PHS, BESS, \ce{H2} & EVs, heating, \ce{H2} & EVs, heating, \ce{H2} \\
\cmidrule{1-10}
Further REMIND-PyPSA developments & REMIND & Regional PyPSA models & EU, China, Globally; 2030-2100 & See above + integration with sector models & See above + electricity trade & Depending on complexity & See above + V2G + industry & See above + e-trucks + industry & See above + e-trucks + industry \\
\bottomrule
\end{tabularx}
\begin{tablenotes}
\scriptsize
\item This table only includes publications that couple two established models with a process-based energy system in the long-term model and explicit data transfer to the ESM. Abbreviations: BESS = Battery Energy Storage System, CAES = compressed air energy storage, CCS = carbon capture and storage, CFs = capacity factors, CSP = concentrated solar power, EVs = electric vehicles, \ce{H2} = hydrogen (electrolysis), PHS = pumped hydro storage, RH = rolling horizon, V2G = vehicle to grid, VRE = variable renewable energy.
\item[a] Costs refers to all kinds of techno-economic parameters, including capital costs, variable costs and \ce{CO2} prices.
\item[b] Backup refers to different parametrisation of required dispatchable capacity.
\item[c] Sectors for which dedicated demand profiles are included in the ESM. This is a necessary condition for modelling demand-side flexibility.
\end{tablenotes}
\end{threeparttable}
\end{table}
\end{landscape}

\subsection{Contribution of this study}

In this study, we extend the state-of-the-art by coupling the IAM REMIND with the ESM PyPSA-Eur through a bi-directional, price-based, and iterative soft coupling. This leverages the complementary strengths of both models and bridges scales to enable the joint optimisation of long-term cross-sectoral mitigation pathways and short-term power system operation (Figure \ref{fig:strengths}). We demonstrate the coupling for a German climate neutrality by 2045 scenario. Specifically, we introduce the following key novelties:

\begin{enumerate}
\item \textbf{Comprehensive bi-directional coupling}: The coupling covers every REMIND timestep, not just a single year or selected milestone years, in a fully automated iterative process, exchanging the complete set of parameters required to harmonise both models\parencite{gongBidirectionalCouplingLongterm2023} and thereby ensuring a fully consistent representation of power system dynamics from PyPSA-Eur in REMIND.
\item \textbf{Sector coupling}: Going beyond aggregated demand, we model the full demand-side evolution by coupling sectoral electricity demands for electric vehicles (EVs), heat pumps, resistive heating, and hydrogen end-use. For example, this captures the power system effects of increased winter peak loads due to rising heat pumps adoption.
\item \textbf{Demand flexibility}: We include endogenous demand-side management for EVs, heat pumps, resistive heating and electrolysers, building on the PyPSA-Eur approach extended with sector-specific data from REMIND and detailed sector models. For the first time, this enables IAM scenarios that incorporate the impact of demand-side flexibility.
\item \textbf{Full price-based coupling}: We integrate price signals from fundamental power system economics through supply-side market values (per technology) and demand-side electricity prices (per sector). REMIND's investment decision therefore embeds power system effects that unfold on hourly time scales into decade-long planning.
\item \textbf{Spatio-temporal resolution}: We use PyPSA-Eur's spatio-temporal capabilities, including renewable profiles, infrastructure data, and optimal grid expansion, with hourly resolution and full foresight across the entire year. This captures seasonal and diurnal balancing and enables the harmonisation of generation capacities despite different spatial resolution.
\end{enumerate}

The remainder of this paper is structured as follows. Section \ref{sec:models} describes both models with example studies. Section \ref{sec:coupling} introduces the bi-directional soft coupling interface, including all variables iteratively exchanged between both models. Section \ref{sec:results} showcases scenario results that reach climate neutrality in Germany by 2045, with and without demand-side flexibility. Section \ref{sec:conclusion} describes limitations of the coupling, identifies avenues for future research, and concludes.

\section{Model description}
\label{sec:models}

\subsection{REMIND}

\begin{figure}
\centering
\includegraphics[width=\textwidth]{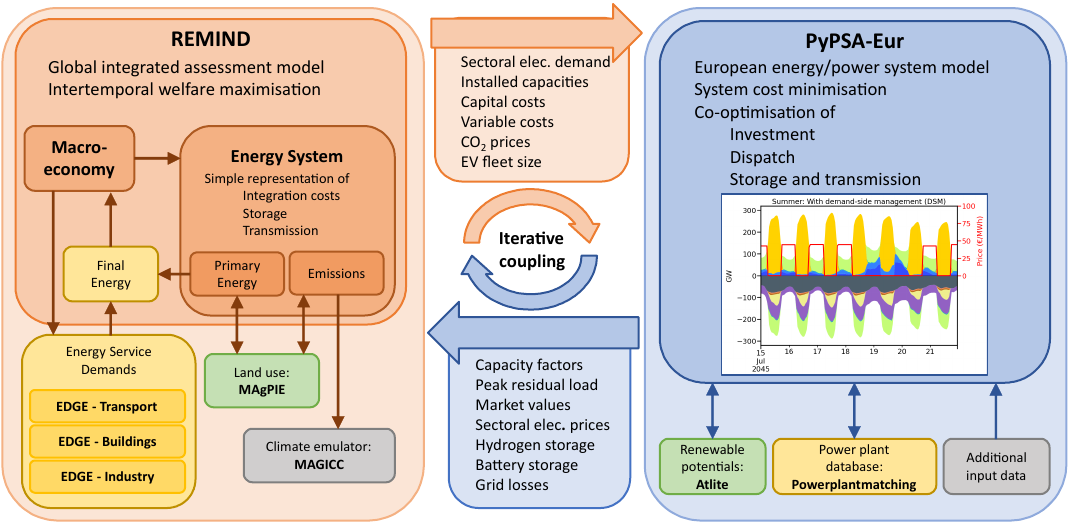}
\caption{\textbf{Bi-directional iterative coupling of REMIND and PyPSA-Eur.} REMIND is a global integrated assessment model that maximises intertemporal welfare and includes a wide portfolio of energy transformation technologies from primary via secondary to final energy with a simple representation of integration costs, storage, and transmission for the power system. REMIND is coupled to dedicated models that provide energy service demands, as well as optionally to the land use model MAgPIE and the climate emulator MAGICC. PyPSA-Eur is a European macro energy system model that minimises annualised system costs by co-optimising investment, dispatch, storage and transmission. Among other data sources, PyPSA-Eur retrieves renewable potentials from the atlite package and power plants from the powerplantmatching package.}
\label{fig:coupling}
\end{figure}

REMIND (REgional Model of INvestments and Development) is a global IAM that links the economy, climate and energy system to investigate self-consistent transformation pathways for climate change mitigation scenarios\parencite{baumstarkREMIND2TransformationInnovation2021}. REMIND explores a wide spectrum of possible futures, linked to technological progress, socioeconomic trends and policy decisions. Assuming perfect foresight, the model uses nonlinear optimisation to maximise intertemporal welfare until 2100 with 5-year time steps until 2060 and 10-year timesteps afterwards, subject to emissions constraints. REMIND currently includes up to 21 world regions, with higher detail in the European Union. It hard-links a macroeconomic Ramsey-type growth model to a detailed representation of the energy system and includes interfaces with the land-use model MAgPIE\parencite{dietrichMAgPIE4Modular2019}, the climate emulator MAGICC\parencite{meinshausenEmulatingCoupledAtmosphereocean2011} and sector-specific EDGE models that provide energy service demands for transport\parencite{rottoliCouplingDetailedTransport2021}, industry\parencite{pehlModellingLongtermIndustry2024}, and buildings\parencite{levesqueHowMuchEnergy2018} (Figure \ref{fig:coupling}). REMIND supports both cost-effectiveness analysis through emissions constraints as well as cost-benefit analysis through macroeconomic damage functions. Via cost penalties for rapid capacity scale-up, REMIND represents inertia in technology transitions, leading to ambitious yet plausible long-term transformation pathways that ensure near-term realism\parencite{weigmannValidationClimateMitigation2025}.

The energy system in REMIND includes a detailed representation of primary, secondary, and final energy carriers, with more than 50 energy conversion technologies\parencite{baumstarkREMIND2TransformationInnovation2021}. Technologies compete based on cost, efficiency and emissions and are represented with full vintage tracking. The model endogenously includes nonlinear technological learning for several technologies such as solar PV, wind and electrolysers. Fossil fuel costs are obtained from extraction curves. REMIND models sector-specific energy demands using nested constant elasticity of substitution (CES) production functions, calibrated to end-use projections from the EDGE models (see above). REMIND’s default power system implementation uses a reduced-form approach based on parametrised integration costs for storage, grid expansion and curtailment that rise with increasing shares of wind and solar\parencite{pietzckerUsingSunDecarbonize2014}. While REMIND’s energy system therefore enables an encompassing analysis of cross-sectoral interactions, it lacks the spatio-temporal detail to represent the underlying power system effects explicitly.

REMIND has been used in many studies and model intercomparison exercises. In the Sixth Assessment Report (AR6) of the IPCC\parencite{ipccClimateChange20222022}, REMIND was the model with the highest number of submitted and vetted scenarios\parencite{sognnaesInfluenceIndividualModels2025} and provided two out of five illustrative mitigation pathways\parencite{ludererImpactDecliningRenewable2022,soergelSustainableDevelopmentPathway2021}. REMIND is one of three IAMs that provide scenarios on global climate policy and technology trends for the Network for Greening the Financial System (NGFS), an initiative of over one hundred central banks worldwide\parencite{richtersNGFSClimateScenarios2024}. More recently, REMIND has been used to provide recommendations on the EU's 2040 climate target\parencite{rodrigues2040GreenhouseGas2023} and a full fossil phase-out in the EU\parencite{schreyerNetzeroZerofossilTransforming2025}, while other studies have explored carbon dioxide removal (CDR)\parencite{merfortSeparatingCO2Emission2025}, electrification pathways for China\parencite{gongMultilevelEmissionImpacts2025}, and the role of demand-side strategies\parencite{vanheerdenDemandsideStrategiesEnable2025}.

REMIND is an open-source model, written in GAMS (General Algebraic Modeling System) and solved using CONOPT 3. In this study, we use REMIND v3.5.1, released on 10 July 2025\parencite{ludererREMINDREgionalModel2025} and introduce a new power system module for the coupling to PyPSA-Eur.

\subsection{PyPSA and PyPSA-Eur}

PyPSA (Python for Power Systems Analysis) is a mature and widely adopted open-source framework for modelling and optimising energy systems\parencite{brownPyPSAPythonPower2018}. At its core, it minimises the system costs of generation, transmission (or delivery) and storage for chosen energy carriers. The PyPSA framework is technology-agnostic and supports modelling with high spatiotemporal resolution in capacity expansion and unit commitment modes, for flexible demand, electrical energy flows and operational reserves, making it especially suitable for modelling future power systems with high VRE shares.

PyPSA-Eur is a comprehensive energy system model built using PyPSA that represents the European energy system at high spatio-temporal resolution\parencite{horschPyPSAEurOpenOptimisation2018,brownPyPSAEurSectorCoupledOpen2025}. PyPSA-Eur models demand and supply for electricity, building heat, transport and industry, including infrastructures for electricity\parencite{gotzensPerformingEnergyModelling2019}, gas, hydrogen and \ce{CO2}. Energy demands in the model are exogenous and price-inelastic. Similarly, capital and fuel costs are exogenous and independent of installation and consumption levels, i.e. do not include a supply curve. In this study, we extract the power system from the full energy system model of PyPSA-Eur, while also using sectoral demand profiles and flexibility settings from the full model (Figure \ref{fig:pypsa}). We run PyPSA-Eur in capacity expansion mode with installed capacities provided by REMIND for free (see below), using hourly resolution with perfect foresight for one weather year. Unit commitment is turned off. Our model version is based on PyPSA-Eur v2025.07.0, released on 11 July 2025\parencite{brownPyPSAEurOpenSectorcoupled2025}, and we use the linear solver Gurobi 12.0.2 in Barrier mode without crossover.

PyPSA-Eur has been used, for example, to analyse the potential role of a European hydrogen\parencite{neumannPotentialRoleHydrogen2023} or \ce{CO2} transport network\parencite{hofmannH2CO2Network2025}, to investigate near-optimal solutions\parencite{neumannNearoptimalFeasibleSpace2021,greevenbroekLittleLoseCase2025}, and to explore myopic pathways towards climate targets\parencite{victoriaSpeedTechnologicalTransformations2022}. Recent methodological advancements include an approximation of nonlinear learning curves by piecewise linearisation\parencite{zeyenEndogenousLearningGreen2023} and price-elastic electricity demand\parencite{brownPriceFormationFuel2025}. Further studies have focused on hard-to-abate sectors, analysing optimal biomass usage\parencite{millingerDiversityBiomassUsage2025}, imports of chemical energy carriers to Germany\parencite{hamppImportOptionsChemical2021} or the impact of green hydrogen production standards\parencite{zeyenTemporalRegulationRenewable2024}. The PyPSA framework is also adopted beyond Europe, for example with PyPSA-USA\parencite{tehranchiPyPSAUSAFlexibleOpensource2025}, PyPSA-Korea\parencite{kwakPyPSAKoreaOpensourceEnergy2025} and PyPSA-China\parencite{PyPSAChinaPIK2025}, which is facilitated by the PyPSA-Earth initiative\parencite{parzenPyPSAEarthNewGlobal2023,abdel-khalekPyPSAEarthSectorcoupledGlobal2025}.

\section{Bi-directional iterative soft coupling}
\label{sec:coupling}

This section describes the bi-directional, iterative, and price-based soft coupling of REMIND and PyPSA-Eur. Figure \ref{fig:coupling} provides an overview of both models and all parameters that need to be exchanged for model harmonisation. Section \ref{subsec:remind2pypsa} lists all parameters that are transferred from REMIND to PyPSA-Eur, while section \ref{subsec:pypsa2remind} conversely lists all parameters that are transferred from PyPSA-Eur to REMIND. Section \ref{subsec:iterative} describes the bi-directional iterative coupling approach and the convergence criteria.

\subsection{REMIND to PyPSA-Eur}
\label{subsec:remind2pypsa}

\begin{figure}
\centering
\includegraphics[width=\textwidth]{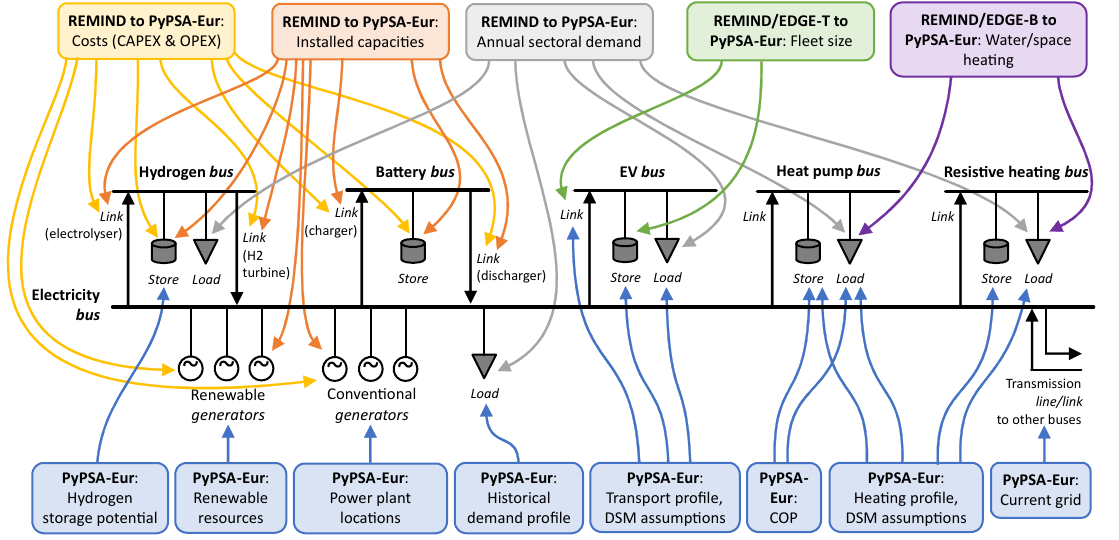}
\caption{\textbf{Illustration of customised PyPSA-Eur model structure with REMIND input data (top) and PyPSA-Eur input data (bottom).} For an explanation of the basic model components (buses, loads, generators, links, stores) please see \parencite{brownPyPSAPythonPower2018}. This is a simplified diagram that only shows a single node. Horizontal lines show buses, to which generators, stores and loads are attached. Links connect the hydrogen, battery, EV, heat pump and resistive heating buses to the main electricity bus. REMIND provides all costs, installed capacities, and annual sectoral demands to PyPSA-Eur. PyPSA-Eur provides data for hydrogen storage potential, renewable resources (potential and availability), power plant locations, sectoral demand profiles and demand-side management (DSM) assumptions. For heat pumps, PyPSA-Eur provides the temperature-dependent, and thus time-dependent, coefficient of performance (COP). Thermal energy storage for heat pumps and resistive heating is implemented as an equivalent electricity storage.}
\label{fig:pypsa}
\end{figure}

REMIND provides the following parameters to PyPSA-Eur. Figure \ref{fig:pypsa} shows an illustration of the customised PyPSA-Eur model structure, indicating for which model components these parameters are used.

\subsubsection{Sectoral electricity demand and flexibility}

REMIND provides annual electricity demand per sector to PyPSA-Eur, distinguishing (i) electric vehicles (EVs), (ii) heat pumps, (iii) resistive heating, (iv) electrolytic hydrogen for end-use sectors and e-fuel production, and (v) residual demand from other sectors. In PyPSA-Eur, for each sector, these annual demands are downscaled to hourly load profiles, including optional demand-side flexibility (Figure \ref{fig:pypsa}). Flexibility settings of the scenarios are described in section \ref{sec:results}.

For passenger EVs, we build on the default PyPSA-Eur implementation, including an hourly load profile, an availability profile for the charger, and an optional storage that represents the total capacity of all EV batteries for demand-side management. The fleet size and the aggregated electricity demand are included from the dedicated EDGE-Transport model, which is also soft-coupled with REMIND\parencite{rottoliCouplingDetailedTransport2021}. We use the fleet’s annual electricity demand to scale the hourly load profile. For simplicity, we do not model bi-directional charging via vehicle-to-grid (V2G). In the future, electric trucks and V2G will be modelled explicitly.

For heat pumps and resistive heating, we use PyPSA-Eur's heating profiles that are based on a daily heating profile scaled by degree days. We enhance this approach by splitting total heating demand into water heating and space heating, based on input data from the EDGE-Buildings model used as part of REMIND. This leads to a baseline demand for water heating, also in summer. For heat pumps, we incorporate the temperature-dependent coefficient of performance (COP) from PyPSA-Eur, leading to reduced efficiency and therefore increased electricity consumption during colder periods. Demand-side management can be modelled through an optional thermal storage with a fixed E/P ratio, implemented as an equivalent electricity storage.

In PyPSA-Eur, hydrogen is produced through electrolysis for both long-term storage and to meet additional demand from end-use sectors in REMIND. For the latter, REMIND provides annual electrolytic hydrogen demand from various applications such as direct reduction of iron for the steel industry, provision of industrial heat, or e-fuel production. We assume a constant hydrogen demand profile, which is spatially distributed according to PyPSA-Eur's electrical load distribution. Flexibility is incorporated through underground cavern storage potentials from PyPSA-Eur.

The residual electricity demand in REMIND encompasses electricity consumption for (i) industry, (ii) road freight transport, and (iii) non-heating demand in buildings. These sub-sectors are aggregated and downscaled to hourly profiles using the default historical load profile in PyPSA-Eur. In the future, these end-uses will also be modelled explicitly.

\subsubsection{Installed capacities}

For each time step, REMIND provides installed capacities to PyPSA-Eur for free in order to ensure realistic transition pathways rather than instantaneous expansion of renewables. Capacities can be provided for generation and storage technologies (Figure \ref{fig:pypsa}). By default, in each time step, we transfer REMIND’s pre-investment capacity to PyPSA-Eur, defined as the capacity from the previous time step minus retired capacity. PyPSA-Eur then needs to further expand capacities in order to meet demand, avoiding the forced deployment of capacities that may be sub-optimal on hourly time scales.

For generation technologies, we harmonise REMIND’s capacities with PyPSA-Eur’s power plant database to preserve locational information while ensuring harmonisation. First, for each year, we remove power plants that are planned to be decommissioned by that year. Second, we compare REMIND’s capacities with this filtered database for each technology. If REMIND’s capacity is smaller than the database total, we scale down the database capacities, equivalent to early retirement. If REMIND’s capacity is larger than the database total, the database remains unchanged and PyPSA-Eur endogenously determines optimal locations for the additional free capacity.

For battery and hydrogen storage technologies, PyPSA-Eur can freely decide where to install the capacities provided by REMIND, respecting hydrogen underground storage potentials. Installed capacities can be provided for all storage components. However, this feature is currently deactivated as it increased solving time with no discernible impact on results.

\subsubsection{Capital cost components}

For all generation and storage technologies, REMIND provides all cost components required to calculate annualised capital costs in PyPSA-Eur. These include specific capital costs and fixed operation and maintenance (FOM) costs, which are annualised given lifetimes and interest rates, both of which are also provided from REMIND. Interest rates are determined from the macroeconomic module in REMIND.

\subsubsection{Marginal cost components}

For all generation and storage technologies, REMIND provides all cost components required to calculate marginal costs in PyPSA-Eur. These include fuel costs, variable operation and maintenance (VOM) costs, efficiencies, as well as \ce{CO2} prices and \ce{CO2} intensities. In REMIND, \ce{CO2} prices are endogenous and are iteratively adjusted to meet defined climate targets.

\subsubsection{Special case: Hydropower}

To harmonise hydropower, we adjust the inflow time series in PyPSA-Eur such that it meets REMIND’s capacity factor. In the future, when extending the model coupling to regions with a substantial share of hydropower, this approach will be revisited. Note that this does not concern pumped hydro storage (PHS), which is a storage technology without any inflow.

\subsection{PyPSA-Eur to REMIND}
\label{subsec:pypsa2remind}

Using the input from REMIND, we solve 14 PyPSA-Eur networks in parallel, one for each time step of REMIND. When all networks are solved, we collect and aggregate the following parameters and pass them back to REMIND.

\subsubsection{Capacity factors}

PyPSA-Eur optimises the investment and utilisation of all generation and storage technologies across the entire year, given free installed capacity from REMIND. We extract annual capacity factors for all generation and storage technologies, which we transfer to REMIND. In contrast to PyPSA-Eur, due to its inter-decadal foresight, REMIND incorporates the development of capacity factors over the full technological lifetime into its investment decision.

\subsubsection{Backup capacity for peak residual load}

To determine the required level of dispatchable backup capacity, we extract the residual peak load from PyPSA-Eur. This represents the highest load across all hours that must be supplied by dispatchable generators, after subtracting variable renewable energy (VRE) generation and any storage supply. REMIND then endogenously determines the optimal mix of dispatchable technologies.

\subsubsection{Market values (supply side)}

\begin{figure}
\centering
\includegraphics[width=\textwidth]{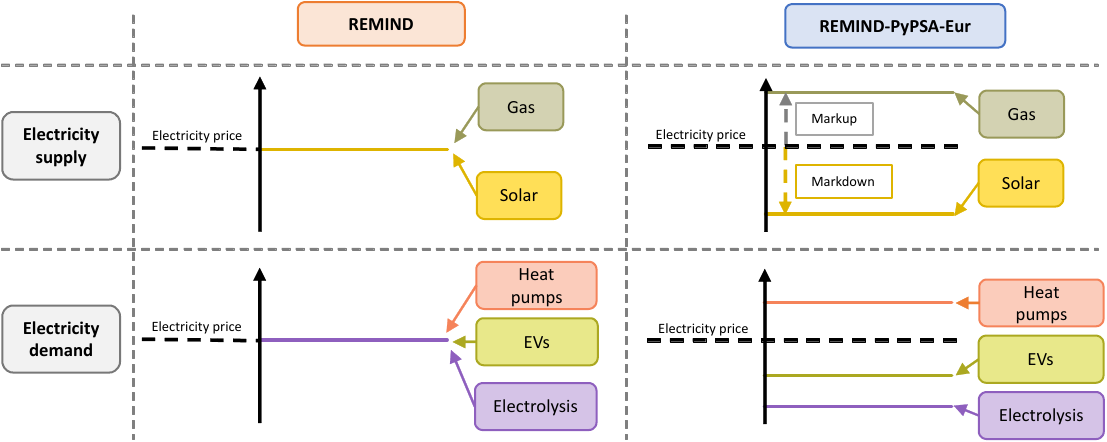}
\caption{\textbf{Illustration of price-based coupling based on market values (supply side) and sectoral electricity prices (demand side).} The first column shows the current modelling in REMIND, where all generation technologies see the same electricity price and therefore the same market value (supply side), while all end-use sectors pay the same electricity price (demand side). The second column shows the new modelling in REMIND-PyPSA-Eur, where generation technologies see different electricity prices, depending on the interactions of their temporal profile and flexibility in PyPSA-Eur, which are passed to REMIND as technology-specific markups and markdowns (supply side). Similarly, end-use sectors pay different electricity prices, also depending on their temporal profile and their potential for demand-side flexibility (demand side).}
\label{fig:markups}
\end{figure}

As a key metric of power system economics, we obtain market values for all generation technologies from PyPSA-Eur. These represent the relative prices that generators receive for each unit of electricity produced, reflecting the value of flexibility on the supply side\parencite{bottgerWholesaleElectricityPrices2022,hirthMarketValueVariable2013}. For example, gas generator will typically achieve above-average, and solar PV below-average market values (Figure \ref{fig:markups}). In REMIND, market values act as price signals that influence long-term investment decisions, thereby representing integration costs and flexibility benefits of high-VRE power systems on the supply side. This is a key innovation compared to IAMs that typically assume uniform electricity pricing across all generation technologies and end-use sectors.

\subsubsection{Sectoral electricity prices (demand side)}

The demand-side equivalent to supply-side market values are sector-specific electricity prices, which capture the relative price paid by different end-use sectors, reflecting their temporal demand profile and flexibility potential. We extract average electricity prices for electric vehicles, heat pumps, resistive heating, electrolysis, and other electricity demand. Sectors using electricity during high-price periods, such as heat pumps that have to operate during winter peak demand, face higher average prices, whereas flexible sectors that can shift their consumption to low-price periods, such as electrolysers, benefit from below-average electricity prices (Figure \ref{fig:markups}). Similar to market values, these relative prices influence the investment decision in REMIND on the demand side, representing the first implementation of sectoral electricity pricing in IAM-ESM coupling.

\subsubsection{Hydrogen and battery storage}

PyPSA-Eur optimises the investment and dispatch of hydrogen and battery storage technologies with perfect foresight for one year, covering daily and seasonal balancing. We transfer all parameters to REMIND that are required to fully harmonise the annual electricity balance between both models: (i) capacity factors for electrolysers, hydrogen turbines, battery chargers and dischargers, (ii) annual electricity generation from hydrogen turbines and battery dischargers, and (iii) required capacities of battery storage and hydrogen underground storage. For consistency with PyPSA-Eur, we require that only electrolytic hydrogen can be used in hydrogen turbines in REMIND. Round-trip losses are automatically harmonised as efficiencies are transferred from REMIND to PyPSA-Eur (see above).

\subsubsection{Grid losses}

PyPSA-Eur calculates optimal transmission expansion and optimised power flows, accounting for the spatial distribution of generation and demand. Grid transmission losses are approximated\parencite{neumannAssessmentsLinearPower2022} and subtracted from REMIND's electricity balance equation.

\subsection{Iterative coupling}
\label{subsec:iterative}

\begin{figure}
\centering
\includegraphics[width=\textwidth]{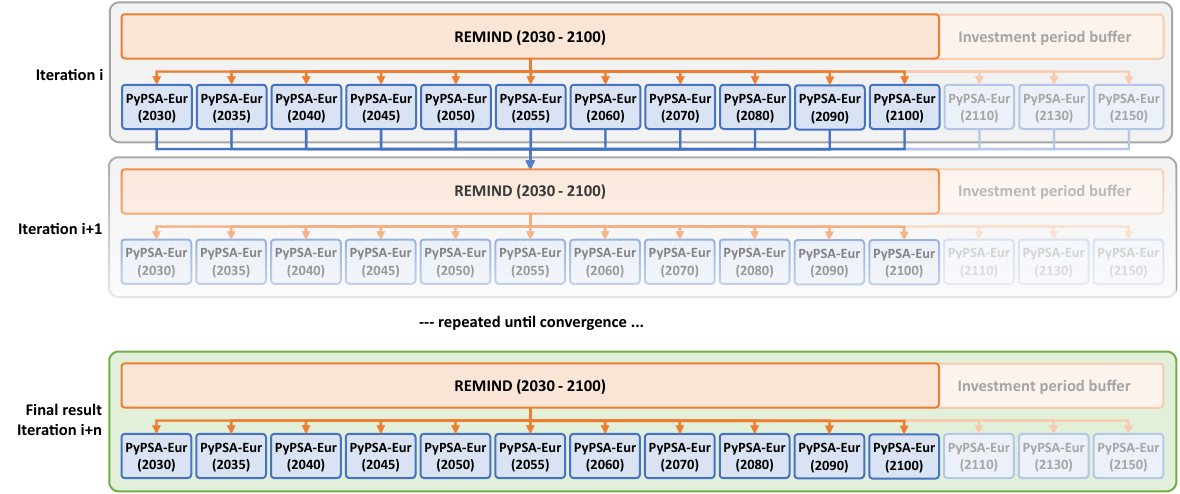}
\caption{\textbf{Iterative workflow of coupling REMIND and PyPSA-Eur.} REMIND runs with full foresight from 2030 until 2100, maximising intertemporal welfare. To avoid end-of-period effects, REMIND also incorporates additional time steps until 2150 as an approximation of a transversality condition. For full consistency, these additional time steps are also coupled to PyPSA-Eur. The coupling to PyPSA-Eur is started in iteration i and then runs iteratively until multiple convergence criteria are met. For each REMIND iteration, 14 PyPSA-Eur networks have to be solved, limiting spatio-temporal detail due to numerical complexity.}
\label{fig:iterative}
\end{figure}

The iterative coupling process ensures convergence between REMIND's long-term optimisation and PyPSA-Eur's detailed power system analysis through a fully automated workflow (Figure \ref{fig:iterative}). Note that REMIND maximises intertemporal welfare with perfect foresight from 2030 to 2100. To avoid end-of-period effects, REMIND uses additional time steps until 2150 as an investment horizon buffer. For full consistency, we also couple these additional time steps. Starting in iteration i, PyPSA-Eur runs for every REMIND time step at hourly resolution with perfect foresight across one year, solving 14 networks in parallel, based on capacity expansion modelling with free installed capacities from REMIND (see above).

The coupling continues iteratively until all parameters transferred from PyPSA-Eur to REMIND reach sufficient convergence. We define convergence as achieved when the relative change in all imported parameters remains below 5\% for four consecutive iterations, balancing accuracy with the computational cost of additional iterations. Better convergence than 5\% was not possible given structural differences between the models that cannot be eliminated through parameter harmonisation, such as different spatial resolution, different modelling foresight, and different objective functions.

To improve convergence and avoid oscillations during the iterative process, we implement anticipation factors that help REMIND anticipate how investment decisions will affect the power system in PyPSA-Eur. These heuristically determined factors are currently applied to capacity factors and market values, building on the work of \textcite{gongBidirectionalCouplingLongterm2023}. For example, when REMIND invests in a technology, the anticipation factor signals that as a result the capacity factor will decrease. In the joint optimum, these anticipation factors vanish and therefore only affect the convergence behaviour, but not the configuration of the equilibrium solution. Future work will focus on determining these factors automatically. 

\section{Scenario results for German climate neutrality by 2045}
\label{sec:results}

This section demonstrates the REMIND-PyPSA-Eur coupling for a Germany climate neutrality target by 2045, distinguishing two illustrative scenarios: (i) with demand-side management, the DSM scenario, and (ii) without demand-side management, the NoDSM scenario. Table \ref{tab:scenarios} provides an overview of the scenario settings for both models. Scenario settings in REMIND are the same in both scenarios.

\begin{table}
\footnotesize
\caption{Scenario settings with and without demand-side management (DSM and NoDSM scenario)}
\label{tab:scenarios}
\begin{tabularx}{\linewidth}{L{1.5cm}L{1.8cm}XX}
\toprule
& & \multicolumn{2}{c}{\textbf{Scenario}} \\
\cmidrule{3-4}
& & \textbf{With demand-side management\newline(\textit{DSM} scenario)} & \textbf{Without demand-side management \newline(\textit{NoDSM} scenario)} \\
\midrule
\multicolumn{4}{l}{\textbf{\textit{REMIND scenario settings}}} \\
\midrule
& General & \multicolumn{2}{L{11cm}}{2°C-compatible scenario (1000 Gt\ce{CO2} peak budget from 2020), based on SSP2, electrolysis pays no grid fees and taxes} \\
\cmidrule{2-4}
& Germany & \multicolumn{2}{L{11cm}}{2030 target:  440 Mt\ce{CO2}eq/a total GHG emissions excl. land-use change \newline 2045 target: GHG-neutrality \newline Biomass: 1.5 EJ/a bioenergy potential, no bioenergy imports \newline Carbon Capture and Storage: geological storage injection limited to 55 Mt\ce{CO2}/a
} \\
\midrule
\multicolumn{4}{l}{\textbf{\textit{PyPSA-Eur scenario settings for electricity demand sectors}}} \\
\midrule
Electrolysis & Demand & Constant demand profile & Constant demand profile \\
\cmidrule{2-4}
& DSM settings & 10\% minimum load, no ramping constraints & 30\% minimum load, ±5\%/hour maximum ramping \\
\cmidrule{1-4}
Passenger EVs & Demand & Hourly profile & Hourly profile (time-shifted to emulate charging profile) \\
\cmidrule{2-4}
& DSM settings & 60 kWh/car (fleet size from REMIND/EDGE-T), 80\% min. charge at 7AM, 50\% participation & - \\
\cmidrule{1-4}
Heat pumps & Demand & Hourly profile based on degree days and COP & Hourly profile based on degree days and COP \\
\cmidrule{2-4}
& DSM settings & E/P = 3 hours w.r.t peak load, all heat pumps operate flexibly & - \\
\cmidrule{1-4}
Resistive heating & Demand & Hourly profile based on degree days & Hourly profile based on degree days \\
\cmidrule{2-4}
& DSM settings & E/P = 2 hours w.r.t peak load, all resistive heating units operate flexibly & - \\
\cmidrule{1-4}
Other demand & Demand & Hourly historical demand profile & Hourly historical demand profile \\
\cmidrule{2-4}
& DSM settings & - & - \\
\bottomrule
\end{tabularx}
\end{table}

The long-term transformation scenario in REMIND is based on the shared socioeconomic pathways (SSP) middle-of-the-road scenario SSP2 and constrains global cumulative \ce{CO2} emissions from 2020 to a maximum of 1000 Gt\ce{CO2}, which is consistent with limiting global warming below 2°C\parencite{forsterIndicatorsGlobalClimate2023}. For Germany, we impose additional emissions targets in line with the federal climate change act\parencite{BundesKlimaschutzgesetz2024}, reaching the intermediate 2030 target as well as climate neutrality in 2045 (Table \ref{tab:scenarios}). Carbon prices are iteratively adjusted until the respective targets are reached and remain constant after 2045. For Germany, key scenarios settings are based on the Ariadne project\parencite{ludererCostefficientEnergyTransition2025}: Bioenergy production is constrained to 1.5 EJ/a\parencite{thranTechnookonomischeAnalyseUnd2019}, long-term bioenergy imports are disabled due to sustainability concerns, and injection of \ce{CO2} into underground storage is limited to 55 Mt\ce{CO2}/a\parencite{merfortEnergiewendeAufNettoNull2023}.

In PyPSA-Eur, we distinguish different end-use flexibility settings for the DSM and NoDSM scenario across four electricity demand sectors (see Table \ref{tab:scenarios} and Figure \ref{fig:pypsa}). These illustrative scenario settings enable us to examine the impact of short-term flexibility in power system operations on long-term transformation pathways.

For electrolysis, we include additional hydrogen demand from REMIND, e.g. for industrial applications, with a constant demand profile in both scenarios. For the DSM scenario, electrolysers are required to always run at least at 10\% of their nominal capacity, while for the NoDSM scenario we raise this to 30\% and introduce a maximum ramping constraint of 5\%/hour, reflecting technical limitations, increased degradation and efficiency penalties that may make fast electrolyser ramping unattractive, particularly for Alkaline electrolysers\parencite{xiaEfficiencyConsistencyEnhancement2023}, as well as onsite hydrogen production without storage.

For passenger EVs, we use the default hourly road transport demand profile from PyPSA-Eur with a morning and evening peak. For the DSM scenario, we parametrise the size of the aggregated EV battery from the fleet size, assuming a 60-kWh average battery size, which needs to be charged at a minimum of 80\% at 7 AM. We further assume that 50\% of EVs charge flexibly. For the NoDSM scenario, we follow PyPSA-Eur’s default implementation, deriving a simplified charging profile by smoothing and time-shifting the hourly transport demand profile.

For electric heating, we use hourly demand profiles scaled by degree-days. For heat pumps, we incorporate the effect of the temperature-dependent coefficient of performance (COP), further raising electricity demand during cold periods. In the DSM scenario, we assume a thermal energy storage, which also imitates thermal inertia of the buildings, through an E/P ratio of 3 hours for heat pumps and 2 hours for resistive heating with respect to the peak load\parencite{rothPowerSectorBenefits2024,schonigerPotentialDecentralHeat2024}. In the NoDSM scenario, heating demand must be met instantly.

\begin{figure}[htb]
\centering
\includegraphics[width=\textwidth]{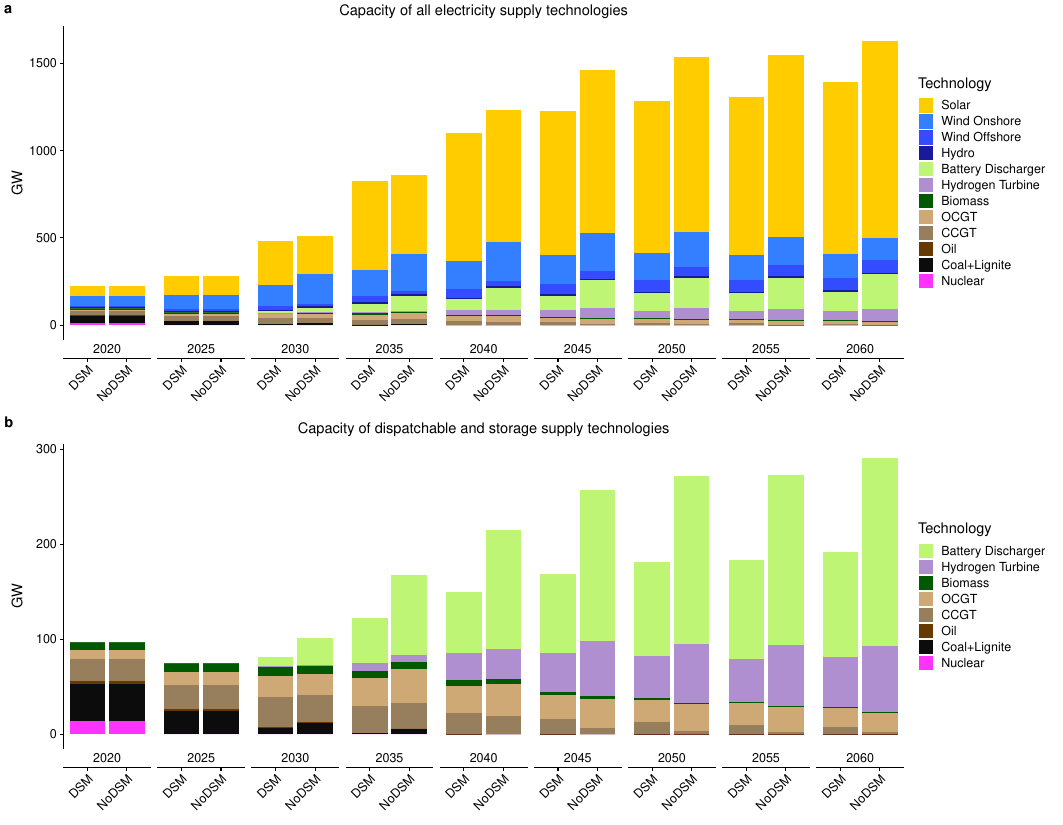}
\caption{\textbf{Optimal capacity in the DSM and NoDSM scenarios until 2060.} \textbf{a}, Optimal capacity of all supply technologies. \textbf{b}, Optimal capacity of fossil dispatchable and storage supply technologies (hydrogen turbines and battery dischargers).}
\label{fig:capacity}
\end{figure}

\subsection{Long-term transition of the electricity system}

Reaching climate neutrality by 2045 requires rapid VRE expansion and storage deployment, with substantial differences between scenarios (Figure \ref{fig:capacity}). The NoDSM scenario demonstrates the infrastructure penalty of inflexible demand: In comparison to the DSM scenario, by 2045 solar PV capacity must increase by 13\% (934 vs. 827 GW), wind by 16\% (263 vs. 225 GW), batteries by 92\% (159 vs. 83 GW), and hydrogen turbines by 41\% (58 vs. 41 GW) compared to the DSM scenario. Only gas turbine capacity decreases by -12\% (37 vs. 42 GW), which however cannot offset the substantial increases of other technologies (Figure \ref{fig:capacity}b). This highlights how DSM enables more efficient capacity utilisation, reducing the need for both VRE overbuilding and large-scale battery storage.

\begin{figure}[hb]
\centering
\includegraphics[width=\textwidth]{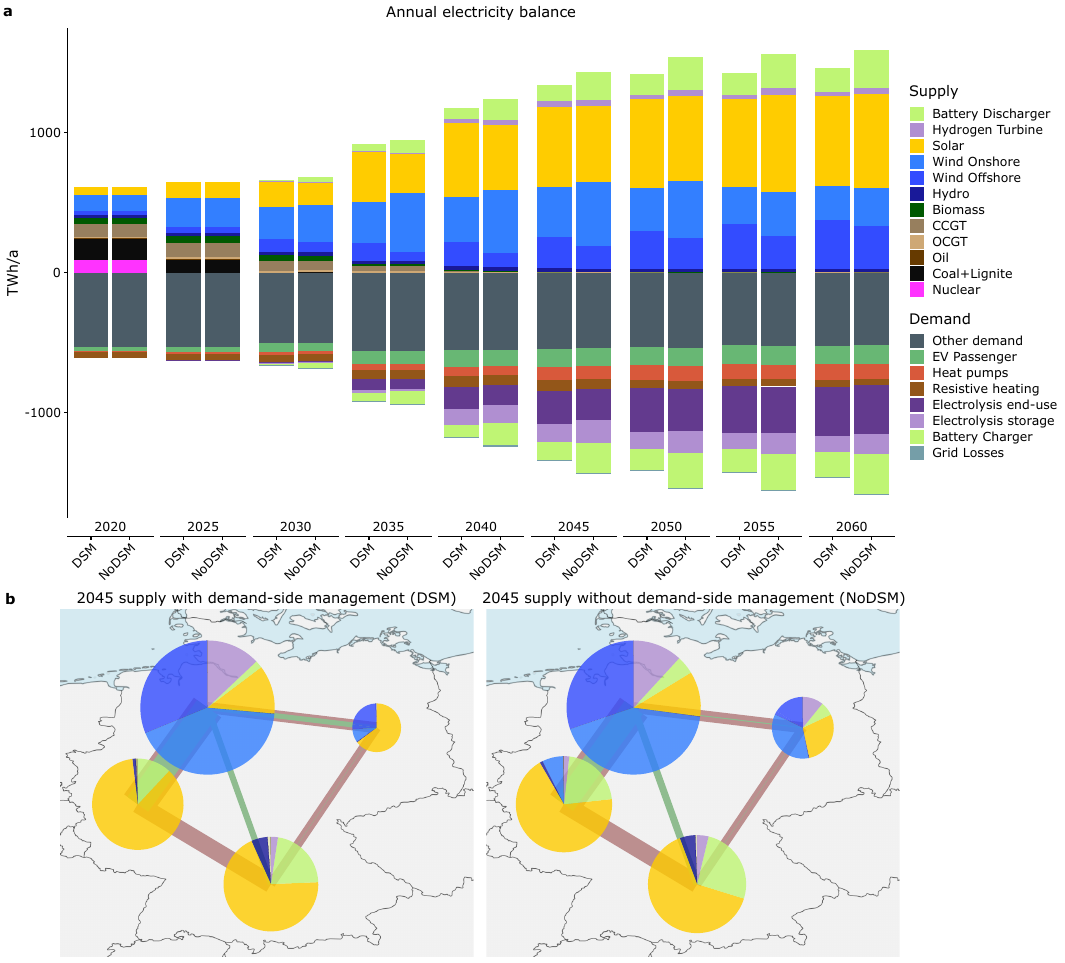}
\caption{\textbf{Annual electricity balance in the DSM and NoDSM scenarios until 2060.} \textbf{a}, Annual electricity balance with positive values showing supply by technology and negative values showing demand by sector. \textbf{b}, Spatial distribution of supply and transmission volumes across 4 nodes (brown is AC transmission, green is DC transmission).}
\label{fig:balance}
\end{figure}

Due to the continued electrification of end-uses, particularly from EVs and heat pumps, as well as the substantially increased demand for electrolytic hydrogen, total electricity demand net of storage increases from 627 TWh/a in 2025 to more than 1000 TWh/a by 2045 in both scenarios (Figure \ref{fig:balance}a). In contrast to installed capacity, the differences in total demand between both scenarios are small with 1078 TWh/a in the DSM scenario and 1054 TWh/a in the NoDSM scenario. In both scenarios, this demand is met almost exclusively by solar PV and wind power, with fossil generation well below 1\%. In response to relative prices from PyPSA-Eur, REMIND adjusts the capacity mix on the supply side and sectoral electricity use on the demand side. On the supply side, in 2045, the DSM scenario has 37\% higher offshore wind power generation in 2045 (223 TWh/a vs. 163 TWh/a) and less solar PV and onshore wind generation, in line with different market values. On the demand side, in 2045, additional ramping flexibility makes hydrogen more attractive, thereby increasing electricity demand for hydrogen end-uses by 5\% in 2045 (233 TWh/a vs. 221 TWh/a), although total electricity demand for electrolysis is larger in the NoDSM scenario, which requires more hydrogen for seasonal storage due to limited demand-side flexibility.

The electricity supply in 2045 shows a clear regional distribution, with Northern Germany producing the vast majority of wind power and Southern Germany relying heavily on solar power and batteries (Figure \ref{fig:balance}b). Notably, the NoDSM scenario leads to more wind power generation in Western and Eastern Germany, reflecting the complementarity between wind and solar power, which is more important with limited demand-side flexibility. Since we do not include hydrogen infrastructure, hydrogen turbines are predominantly co-located with hydrogen underground storage in Northern Germany.

\subsection{Daily balancing in summer and winter}

\begin{figure}
\centering
\includegraphics[width=\textwidth]{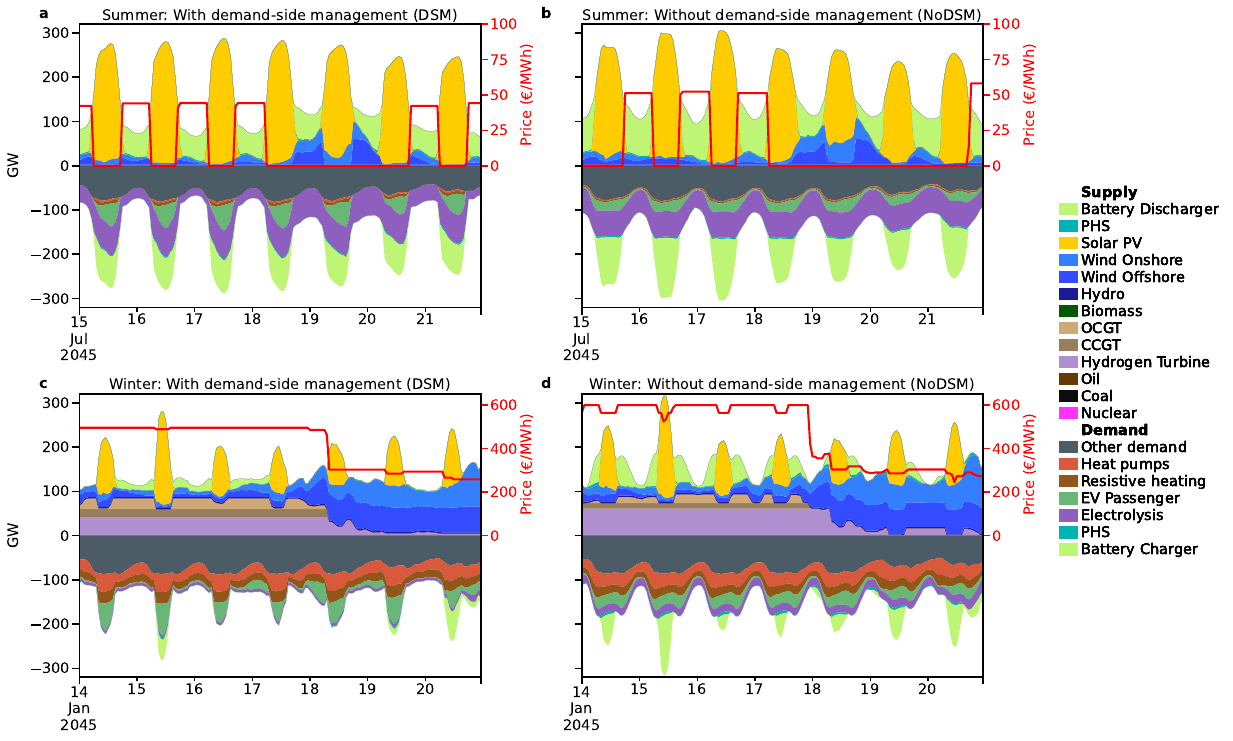}
\caption{\textbf{Hourly electricity balance for a representative week in summer and winter in 2045 for the DSM and the NoDSM scenario.} Positive values show supply from generation technologies and discharging storage technologies, e.g. hydrogen turbines and battery discharging. Negative values show demand by sector and charging storage technologies, e.g. electrolysis and battery charging. The red line indicates hourly prices (weighted by nodal demand) and corresponds to the secondary y-axis on the right.}
\label{fig:hourly}
\end{figure}

The transition to a VRE-dominated power system fundamentally changes electricity supply and demand patterns, with pronounced seasonal and diurnal variations that are critical for system planning. Representative weeks in summer and winter highlight the profound impact of demand-side flexibility in managing VRE variability and integrating electrified demands at the point of climate neutrality in 2045 (Figure \ref{fig:hourly}).

The summer week illustrates pronounced diurnal patterns (Figure \ref{fig:hourly}a-b). In the DSM scenario, substantial solar peaks during daytime hours are effectively absorbed by flexible loads from EVs, electrolysis and batteries, with limited flexibility from electrified water heating (Figure \ref{fig:hourly}a). At night, total electricity demand drops dramatically from more than 250 GW to below 100 GW, with batteries and wind power supplying the reduced load. Electricity prices follow this diurnal pattern closely, reaching zero during peak solar hours and rising to approximately 50 €/MWh at night, except during periods with additional wind generation towards the end of the week. In contrast, the NoDSM scenario shows more dampened diurnal demand variations (Figure \ref{fig:hourly}b). Solar peaks are absorbed by batteries and electrolysis, but cannot utilise the flexibility of EVs. The demand reduction at night is less pronounced, requiring more battery discharging to meet the load. Price variations follow a similar pattern than in the DSM scenario, but are stronger in magnitude.

The winter week represents a renewable-scarce period with limited sunshine and wind during the first four days, highlighting the challenges of maintaining system balance during a mild ``Dunkelflaute'' (Figure \ref{fig:hourly}c-d). Both scenarios show a higher baseload compared to summer due to electrified heating, which is however partially offset by strongly decreased demand from electrolysers. In the DSM scenario, EVs can still make full use of the mid-day solar peak, while heat pumps and resistive heating also shift their demand to a smaller degree. Electricity prices are far higher than in summer, exceeding 500 €/MWh as hydrogen turbines are price-setting and have to recover their capital investment costs within limited operating hours. The NoDSM scenario exhibits similar day-night load variations but faces additional operational constraints as EV and heating demand cannot shift in time and as electrolysis must run at 30\% of its nameplate capacity. Therefore, the mid-day solar peak occurs between two battery discharging periods. Batteries are recharged during the peak, effectively enabling flexibility on similar time scales compared to demand-side management, albeit at additional costs. Electricity prices in the NoDSM scenario are even higher than in the DSM scenario, reflecting the cost of inflexibility.

\subsection{Seasonal balancing across the year}

\begin{figure}
\centering
\includegraphics[width=\textwidth]{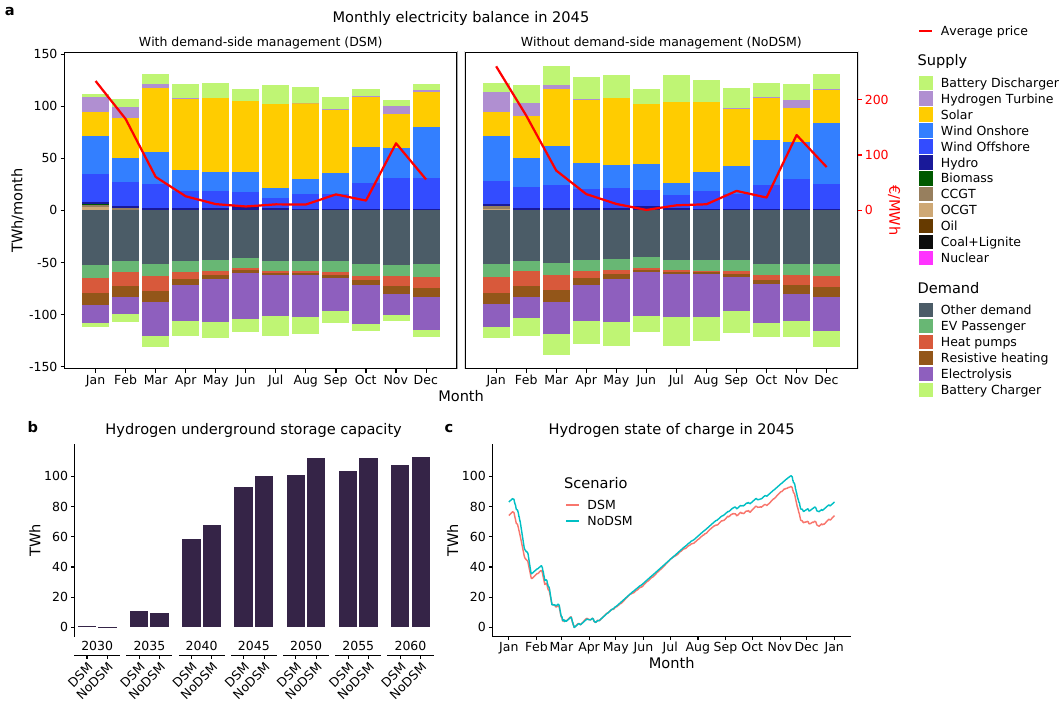}
\caption{\textbf{Monthly electricity balance in 2045 and hydrogen storage in the DSM and NoDSM scenario.} \textbf{a}, Monthly electricity balance and monthly prices in 2045 for both scenarios with positive values showing supply by technology and negative values showing demand by sector. \textbf{b}, Hydrogen underground storage capacity across both scenarios until 2060. \textbf{c}, State of charge of hydrogen underground storage in 2045.}
\label{fig:seasonal}
\end{figure}

The seasonal mismatch of VRE-dominated power systems becomes particularly pronounced in a future climate-neutral energy system with abundant solar electricity generation during summer and high demand for electrified heating during winter (Figure \ref{fig:seasonal}a). In both scenarios, at the point of climate neutrality in 2045, this imbalance is addressed through higher electricity demand for hydrogen production during summer and higher electricity supply through subsequent re-electrification in hydrogen turbines during winter. The DSM scenario can utilise enhanced electrolyser flexibility to produce more hydrogen during summer compared to the NoDSM scenario (44 TWh vs. 41 TWh in June, 7\%), while producing less during winter (18 TWh vs. 23 TWh in January, -21\%). Average monthly electricity prices show strong volatility throughout the year, as prices remain below 20 €/MWh from May to August, but exceed 200 €/MWh in January across both scenarios. As discussed before, winter prices are driven by the high marginal cost of dispatchable generators as well as by scarcity prices of hydrogen turbines that must recover their investment costs. Notably, the substantial wind generation in December leads to a price drop in both scenarios, highlighting the considerable dependence of future electricity prices on renewable resources.

Although demand-side flexibility leads to less deployment of hydrogen turbines (see Figure \ref{fig:capacity}b), the optimal hydrogen underground storage size does not differ substantially between scenarios (Figure \ref{fig:seasonal}b). In both scenarios, storage needs to grows rapidly, reaching reach 93 TWh and 100 TWh in the DSM and NoDSM scenario by 2045, respectively. Hydrogen storage fills up continuously during spring and summer, tapping into abundant and low-cost renewable supply, and depletes during autumn and winter as hydrogen turbines provide dispatchable electricity (Figure \ref{fig:seasonal}c). Note that this storage capacity is not only used for seasonal storage, but also to meet additional demand for hydrogen from end-use sectors in REMIND (see Figure \ref{fig:pypsa} and Figure \ref{fig:balance}).

\subsection{Electricity economics and relative prices}

\begin{figure}
\centering
\includegraphics[width=\textwidth]{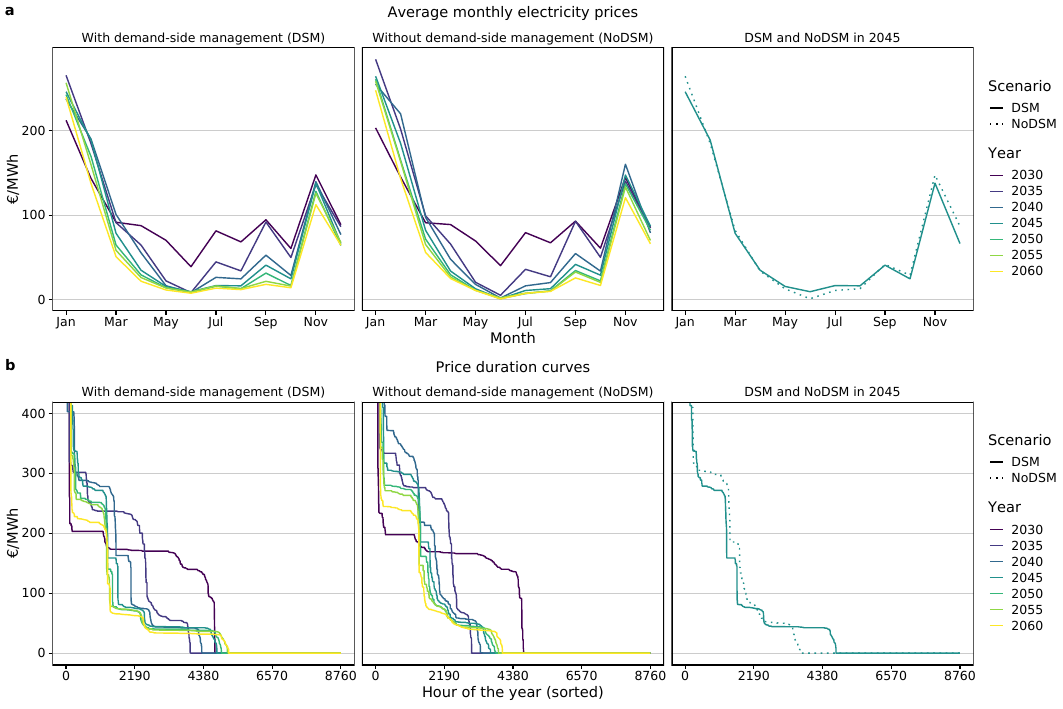}
\caption{\textbf{Average monthly prices and price duration curves for the DSM and NoDSM scenario until 2060.} \textbf{a}, Average monthly electricity prices in both scenarios across years until 2060 and scenario comparison for 2045. \textbf{b}, Price duration curves in both scenario across years until 2060 and scenario comparison for 2045.}
\label{fig:prices}
\end{figure}

As shown in the previous section for 2045, electricity prices follow a clear seasonal pattern that corresponds to renewable scarcity. This pattern intensifies over time as the electricity system transitions towards high VRE shares (Figure \ref{fig:prices}a). In 2030, average monthly prices remain within 40-100 €/MWh during spring and summer and only briefly exceed 200 €/MWh in winter. However, by 2045 average monthly prices regularly fall below 20 €/MWh during summer. By 2060 more than half of the months see prices below 30 €/MWh, while prices almost reach 250 €/MWh in January across both scenarios. Demand-side flexibility reduces these seasonal variations, particularly during summer when the DSM scenario enables better utilisation of renewables.

The effect of demand-side flexibility on price volatility is also evident in the price duration curves (Figure \ref{fig:prices}b). In 2045, without demand-side flexibility more than 50\% of all hours show zero prices due to the cost-optimal overbuilding of VRE capacity, which also leads to high curtailment rates. In contrast, zero-price hours can be reduced by demand-side flexibility that helps to absorb surplus renewable generation for seasonal storage or flexible end-use demand. We note that our results indicate a high number of zero-price hours compared to recent research\parencite{geisPriceFormationHighlyRenewable2025}, which likely results from different cost assumptions and limited spatial resolution in our model. Similar to the average monthly prices, the price duration curves also show the increase of volatility over time.

\begin{figure}
\centering
\includegraphics[width=\textwidth]{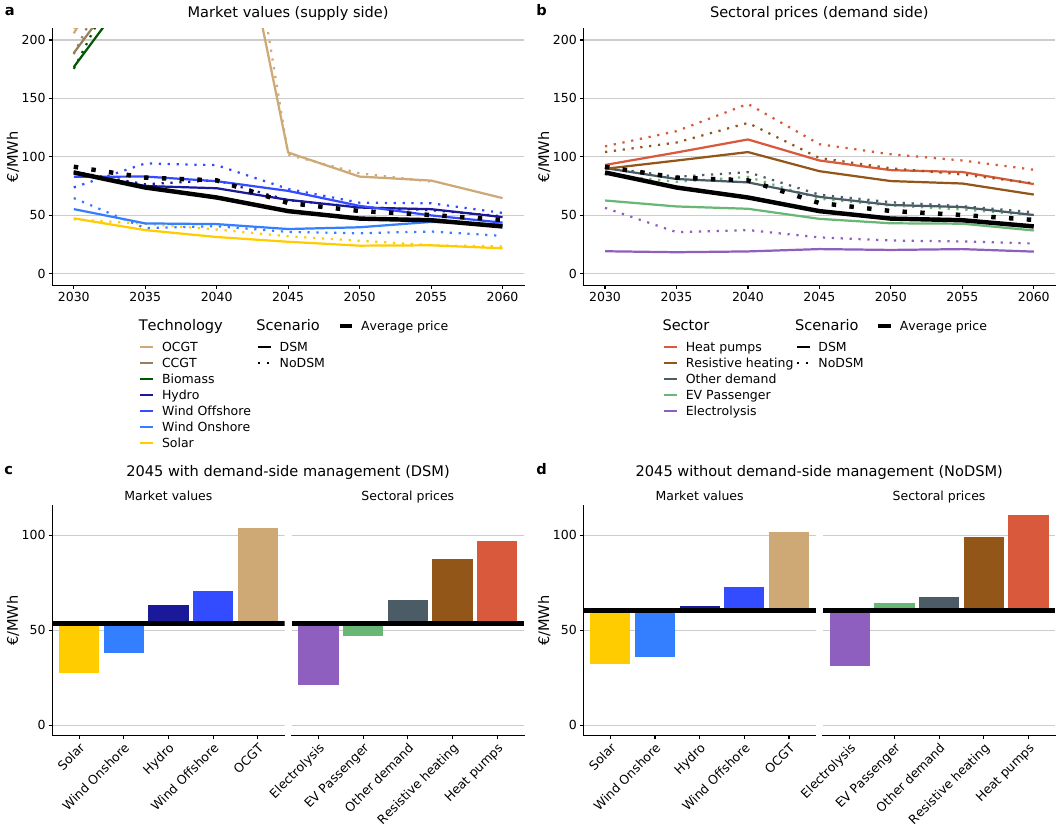}
\caption{\textbf{Market values and sectoral electricity prices for the DSM and NoDSM scenario.} \textbf{a}, Market values on the supply side by scenario, only shown for technologies that supply at least 0.1\% of electricity. \textbf{b}, Sectoral electricity prices on the demand side by scenario. \textbf{c-d}, Average electricity price and markups/markdowns on the supply-side market values and demand-side sectoral prices in 2045 for the DSM scenario (c) and NoDSM scenario (d). Scarcity prices, defined as hourly prices larger than 3 standard deviations above the average, are removed before calculating both market values and sectoral electricity prices.}
\label{fig:marketvalues}
\end{figure}

The pronounced daily and seasonal electricity price fluctuations translate into average price signals, based on fundamental power system economics. On the supply side, market values represent the average revenue generators earn per MWh on the market, weighted by their generation profiles (Figure \ref{fig:marketvalues}a). Conversely, on the demand side, sectoral prices capture the average costs paid by different end-use sectors, weighted by their demand profiles (Figure \ref{fig:marketvalues}b). Market values and sectoral prices are a disaggregation of the average electricity price: the weighted average across all generation technologies or all end-use sectors produces the average electricity price, such that the average price in Figure \ref{fig:marketvalues}a is the same as in Figure \ref{fig:marketvalues}b. In the model coupling, REMIND receives these price signals from PyPSA-Eur, thereby integrating the economic costs of variability and the economic benefits of flexibility into its long-term investment decision. Across all time steps, demand-side flexibility substantially reduces the average wholesale electricity price by up to 12\%. Note that all prices are wholesale prices before taxes and grid fees.

Market values show substantial differences across generation technologies (Figure \ref{fig:marketvalues}a). In both scenarios, dispatchable fossil generators (OCGT, CCGT, biomass) typically produce at market values above 150 €/MWh by supplying electricity during scarcity periods. In contrast, solar PV produces at below-average market values due to considerable mid-day overproduction in both scenarios (27 €/MWh and 32 €/MWh in the DSM and NoDSM scenarios in 2045, respectively). Although onshore wind also experiences below-average market values, offshore wind attains market values at or above the average (around 70 €/MWh in both scenarios in 2045, reflecting the more steady availability of offshore wind. Demand-side flexibility reduces average prices and consequently also leads to lower market values across all generation technologies (Figure \ref{fig:marketvalues}c-d).

Sectoral electricity prices also vary considerably due to temporal demand patterns and flexibility (Figure \ref{fig:marketvalues}b). In both scenarios, electrolysers pay the lowest prices, 21 €/MWh in the DSM scenario and 31 €/MWh in NoDSM scenario in 2045 (before taxes and grid fees), by utilising periods of renewable abundance during summer. For EVs, the average price depends on their flexibility. In the DSM scenario, EVs benefit from flexible charging during solar peaks, paying slightly below average at 47 €/MWh in 2045, while in the NoDSM scenario, they pay an above-average price of 64 €/MWh. Notably, the benefit of flexible charging is most pronounced in 2030 and gradually reduces over time due to competing flexible demands, particularly electrolysis. In stark contrast, electric heating in the buildings sector face the highest average electricity prices as they operate mainly during winter. Although flexible operation reduces their electricity price from 111 €/MWh to 97 €/MWh in 2045, heat pumps still pay nearly twice the average price in both scenarios (Figure \ref{fig:marketvalues}c-d). Notably, the electricity price paid by heat pumps and resistive heating increases until 2040 in both scenarios and declines steadily afterwards, suggesting that the cost of seasonal balancing required to meet electrified heating demand during winter is particularly large in the mid-term.

\subsection{Demand-side flexibility across sectors}

\begin{figure}[htb]
\centering
\includegraphics[width=\textwidth]{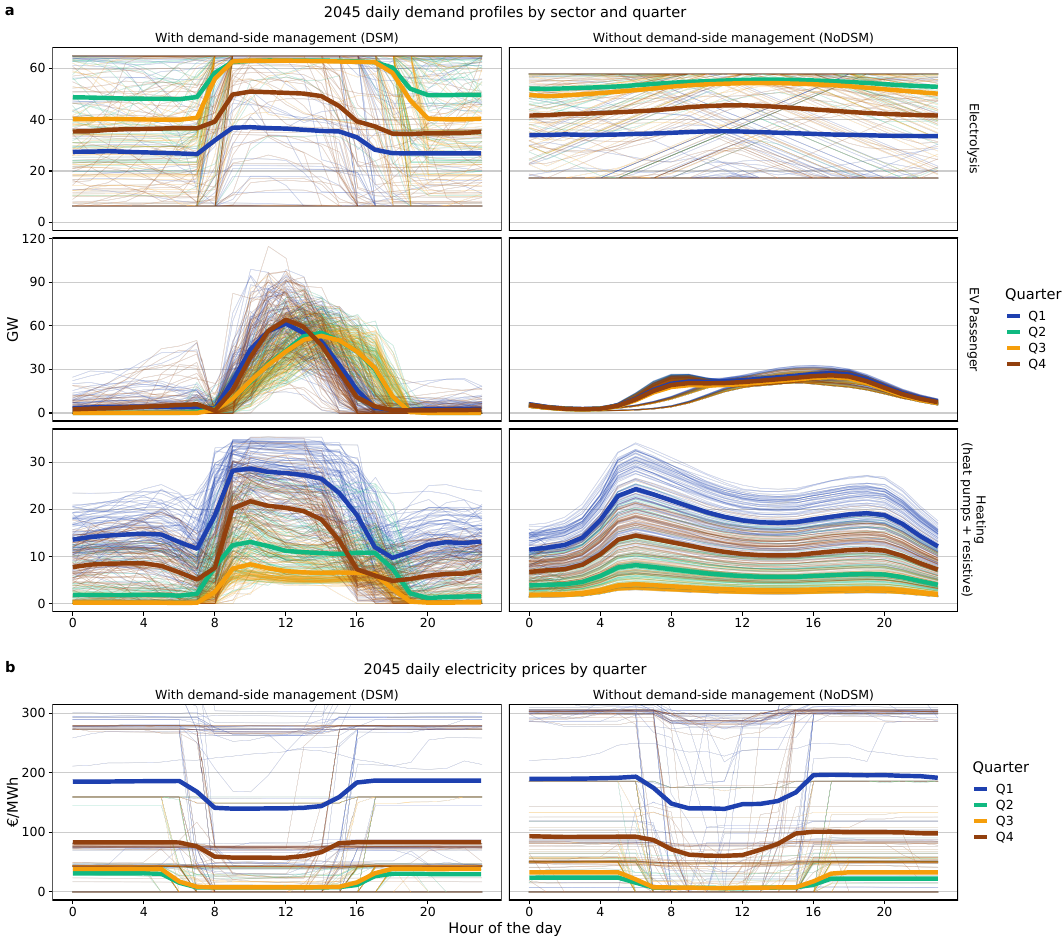}
\caption{\textbf{Daily demand profiles by sector and quarter and daily electricity prices by quarter in 2045 for the DSM and NoDSM scenario.} \textbf{a}, Daily demand profiles by sector (rows) and scenarios (columns) in 2045, distinguished by quarter (colour) with Q1 (Jan-Mar), Q2 (Apr-Jun), Q3 (Jul-Sep), and Q4 (Oct-Dec). Thin lines indicate the daily demand for all 365 days, while thick lines indicate the average for each quarter. Only sectors that include the option for demand-side management are shown, i.e. not the “other demand” category (see Table \ref{tab:scenarios}). \textbf{b}, Daily electricity prices per scenario in 2045 by quarter. In the DSM scenario, sectors shift their demand to hours with lower electricity prices, particularly for passenger EVs. In the NoDSM scenario, sectors other than electrolysis cannot shift their demand in time.}
\label{fig:profiles}
\end{figure}

The variation in sectoral electricity prices is a direct result of their demand profile and their demand-side flexibility potential. Across both scenarios, hourly demand profiles per day illustrate how the modelled sectors shift their demand in time (Figure \ref{fig:profiles}a) in response to hourly prices (Figure \ref{fig:profiles}b), with thin lines indicating daily profiles for all 365 days of the year and thick lines showing average daily profiles by quarter.

Electrolysers show a strong demand response to prices. In both scenarios, electrolysers operate at near-full capacity during midday solar peaks in spring and summer (Q2/Q3), tapping into low prices and filling hydrogen storage for winter. In the less-constrained DSM scenario, daily variations are particularly pronounced, often ramping-up electrolysis quickly when the sun rises and ramping-down just as quickly when the sun sets. In contrast, in the NoDSM scenario, the limited flexibility of electrolysers leads to far less pronounced daily cycling, increasing the average price paid by electrolysers (see Figure \ref{fig:marketvalues}b). Across both scenarios, electrolysers run far less in autumn and winter (Q4/Q1) than in spring and summer (Q2/Q3), due to the seasonality of hydrogen production (see Figure \ref{fig:seasonal}).

Heating demand, shown here as the sum of heat pumps and resistive heating, also show moderate flexibility. Unsurprisingly, across both scenarios, the demand profile is highly seasonal with peak demand from space heating in winter (Q1) and lowest demand from water heating in summer (Q3). Without flexibility, heating demand follows a daily pattern with two modest peaks in the morning and evening, which is an unfavourable profile since prices are typically higher during these times (Figure \ref{fig:profiles}b), resulting in the highest prices of all end-use sectors modelled (see Figure \ref{fig:marketvalues}d). In comparison, the DSM scenario shifts heating demand to mid-day as much as possible, creating a relatively constant demand profile from around 9AM to 2PM in winter and autumn. However, unlike for flexible EV charging, a substantial part of the load still remains at night, resulting in high average prices for heating technologies even with demand-side flexibility (see Figure \ref{fig:marketvalues}c). 

\section{Discussion and conclusion}
\label{sec:conclusion}

This paper presented a bi-directional, iterative, and price-based soft coupling of REMIND and PyPSA-Eur, combining the complementary strengths of both models and resolving a fundamental trade-off in energy system modelling between wide scope needed for long-term investment decisions and high spatio-temporal detail needed for short-term power system balancing.

Our approach offers five key innovations that distinguish it from previous coupling efforts. First, the comprehensive iterative coupling harmonises investment decisions and energy balances across both models, creating an energy system modelling framework that bridges inter-decadal investment time scales and intra-annual operation time scales. Second, we explicitly model sector coupling by incorporating evolving demand patterns from electric vehicles, electric heating, and hydrogen end-uses, capturing the power system effects of accelerated electrification and power-to-molecule conversion. Third, we integrate cost-optimal demand-side flexibility across multiple sectors, enabling the first analysis of smart electrification within long-term climate mitigation pathways. Fourth, we use relative prices as a key coupling metric, capturing fundamental power system economics within long-term planning through market values and sectoral prices. Fifth, by incorporating geospatial input data for renewable resources and energy infrastructure, we bridge not only temporal but also spatial scales between both models.

Our scenario analysis for Germany confirms that a sector-coupled energy system with nearly 100\% renewable electricity is technically feasible and cost-effective for reaching climate neutrality. Furthermore, we demonstrate how demand-side flexibility alters optimal electricity systems and influences long-term pathways. While demand-side flexibility enables more efficient capacity utilisation, reducing VRE overbuilding requirements and dispatchable backup needs, inflexible systems need to rely more strongly on battery and hydrogen storage. The DSM scenario achieves system-wide economic benefits with average electricity prices up to 12\% lower than the inflexible NoDSM scenario, demonstrating how smart electrification can help to reduce costs. However, significant sectoral price disparities persist even with demand-side flexibility. Heating applications face electricity prices up to 93\% above average due to unavoidable winter peak loads, while flexible electrolysers access prices up to 78\% below average by shifting demand to periods with abundant cheap renewable supply.

These findings illustrate that policymakers should pursue an electricity market design that supports rather than hampers flexible electrification. This requires an accelerated roll-out of smart meters together with market mechanisms that can provide price signals to consumers to enable demand response across sectors. The pronounced sectoral price differentiation highlights the need for corresponding electricity tariffs that reflect the true hourly cost. However, the substantial price premium for heating applications also points towards political challenges as high electricity prices for some end-users might undermine public support for electrification policies, requiring careful balancing of economic efficiency with political feasibility.

Despite the methodological innovations, several important limitations remain. Most critically, modelling Germany as an electrical island neglects the substantial benefits of European electricity trade for balancing VRE variability and seasonal storage requirements, likely overestimating domestic storage and backup capacity needs\parencite{brownSynergiesSectorCoupling2018,schlachtbergerBenefitsCooperationHighly2017}. The use of a single weather year cannot capture inter-annual meteorological variability and uncertainty\parencite{schmidtLongdurationStorageWeather2025}, potentially leading to an underestimation of backup capacity requirements\parencite{gotskeDesigningSectorcoupledEuropean2024,rugglesPlanningReliableWind2024}. Similarly, we do not account for the impact of climate change on the supply and demand of future energy systems\parencite{vandermostImpactsFutureChanges2025}. The current implementation also excludes hydrogen infrastructure and does not model heating networks, which could offer the flexibility to switch heating sources depending on the electricity price, as well as unlocking long-duration aquifer or pit thermal energy storage.

The coupling framework opens numerous avenues for future research. While we are already working on a geographic expansion to the full European electricity system, further extensions to China and other regions are also in development, potentially using the PyPSA-Earth framework \parencite{parzenPyPSAEarthNewGlobal2023,abdel-khalekPyPSAEarthSectorcoupledGlobal2025}. Methodological improvements should focus on endogenous determination of anticipation factors, integration of reserve margins, and the implementation of price-elastic demand to improve the modelling of price formation and make results less dependent on single weather years\parencite{brownPriceFormationFuel2025}. Further integration with sector models for transport\parencite{rottoliCouplingDetailedTransport2021} and buildings\parencite{hasseBrickBuildingSector2025} that are already linked to REMIND could lead to an improved understanding of the effect of sectoral electricity prices on demand-side transformations, for example for road freight transport and heat pump deployment. The demand-side flexibility parametrisation in PyPSA-Eur could be refined by capturing heterogeneity across end-users, particularly for electric vehicles\parencite{muesselAccurateScalableRepresentation2023} and by incorporating estimates of real-world additional costs of demand-side flexibility.

With electricity positioned to become the dominant energy carrier in future energy systems, policymakers face the critical challenge of optimally integrating variable renewable energy with newly electrified demands in transport, heating, and industry through sector coupling. This challenge is becoming more urgent given the slow deployment of competing mitigation options such as CCS and hydrogen\parencite{kazlouFeasibleDeploymentCarbon2024,odenwellerGreenHydrogenAmbition2025}, recent estimates of limited CCS potential\parencite{giddenPrudentPlanetaryLimit2025}, as well as sustainability concerns around large-scale biomass usage\parencite{ludererEnvironmentalCobenefitsAdverse2019}. As a result, future energy systems may be shaped primarily by the continued acceleration of recent trends: ever-cheaper renewable electricity, rapidly improving battery technologies, and ongoing electrification across sectors. Against this backdrop, the model coupling of REMIND and PyPSA-Eur represents a first step towards an integrated energy system modelling suite that bridges scales and research communities, contributing to a sound scientific evidence base for an increasingly complex, and increasingly electric, future energy system.

\section*{Author contributions (CRediT)}
\setlength{\parindent}{0pt}

\textbf{Adrian Odenweller}: Conceptualization, Methodology, Software, Validation, Formal analysis, Investigation, Data curation, Writing - Original Draft, Writing - Review \& Editing, Visualization

\textbf{Falko Ueckerdt}: Conceptualization, Methodology, Writing - Review \& Editing, Supervision, Project administration, Funding acquisition

\textbf{Johannes Hampp}: Methodology, Software, Validation, Formal analysis, Data curation, Writing - Review \& Editing

\textbf{Ivan Ramirez}: Methodology, Software, Writing - Review \& Editing

\textbf{Felix Schreyer}: Methodology, Software, Validation, Investigation, Writing - Review \& Editing

\textbf{Robin Hasse}: Methodology, Validation, Investigation, Writing - Review \& Editing

\textbf{Jarusch Müßel}: Methodology, Validation, Investigation, Writing - Review \& Editing

\textbf{Chris Chen Gong}: Conceptualization, Methodology, Project administration, Funding acquisition

\textbf{Robert Pietzcker}: Conceptualization, Methodology, Validation, Investigation

\textbf{Tom Brown}: Conceptualization, Methodology, Writing - Review \& Editing

\textbf{Gunnar Luderer}: Conceptualization, Methodology, Validation, Writing - Review \& Editing, Supervision, Project administration, Funding acquisition

\section*{Declaration of competing interest}

None.

\section*{Acknowledgements}

We thank Julian Geis for answering questions related to demand-side management modelling in PyPSA-Eur. We thank Lavinia Baumstark, Falk Benke and Tonn Rüter for technical help with the high-performance computer system at PIK.

\section*{Funding sources}

We gratefully acknowledge funding from the Kopernikus-Projekt Ariadne through the German Federal Ministry of Education and Research (Grant Nos. 03SFK5A and 03SFK5A0-2), the PRISMA project through the European Union’s Horizon Europe research and innovation programme (Grant No. 101081604), the PANDA project supported by Energy Foundation China (Grant No. G-2407-35694), as well as the academic scholarship foundation Villigst (J.M.). and the German Federal Environmental Foundation (A.O.). We also gratefully acknowledge funding from the Ministry of Research, Science and Culture (MWFK) of Land Brandenburg for supporting this project by providing resources on the high-performance computer system at the Potsdam Institute for Climate Impact Research (Grant No. 22-Z105-05/002/001).

\section*{Data availability}

Both REMIND and PyPSA-Eur are open source models. The model code for REMIND, including the new power module realisation developed for this paper, is available at \url{https://github.com/aodenweller/remind/tree/dev/remind_v3.5.1_pypsa} and will be merged with the main REMIND repository, pending checks and adjustments to reporting functions. The customised model code for PyPSA-Eur is available at \url{https://github.com/aodenweller/pypsa-eur/tree/dev/pypsa_v2025.07.0_remind} and will become part of the Potsdam Integrated Assessment Modelling (PIAM) framework, pending documentation and harmonisation with other regional PyPSA models that will also be coupled to REMIND.

\printbibliography

\end{document}